\documentclass[review]{elsarticle}

\makeatletter
\def\ps@pprintTitle{%
 \let\@oddhead\@empty
 \let\@evenhead\@empty
 \def\@oddfoot{\centerline{\thepage}}%
 \let\@evenfoot\@oddfoot}
\makeatother

\usepackage{amsmath}
\usepackage{amssymb}
\usepackage{lscape}
\usepackage{multirow}
\usepackage{color}
\usepackage{natbib}
\usepackage{lineno}

\newcommand*\patchAmsMathEnvironmentForLineno[1]{%
  \expandafter\let\csname old#1\expandafter\endcsname\csname #1\endcsname
  \expandafter\let\csname oldend#1\expandafter\endcsname\csname end#1\endcsname
  \renewenvironment{#1}%
     {\linenomath\csname old#1\endcsname}%
     {\csname oldend#1\endcsname\endlinenomath}}%
\newcommand*\patchBothAmsMathEnvironmentsForLineno[1]{%
  \patchAmsMathEnvironmentForLineno{#1}%
  \patchAmsMathEnvironmentForLineno{#1*}}%
\AtBeginDocument{%
\patchBothAmsMathEnvironmentsForLineno{equation}%
\patchBothAmsMathEnvironmentsForLineno{align}%
\patchBothAmsMathEnvironmentsForLineno{flalign}%
\patchBothAmsMathEnvironmentsForLineno{alignat}%
\patchBothAmsMathEnvironmentsForLineno{gather}%
\patchBothAmsMathEnvironmentsForLineno{multline}%
}

\begin{document}

\begin{frontmatter}

\linespread{1.0}


\title{Multi-rate mass transfer modeling of two-phase flow in highly
  heterogeneous fractured and porous media}


\author[Hannover]{Jan Tecklenburg\corref{cor}}
\cortext[cor]{Corresponding author}
\ead{tecklenburg@hydromech.uni-hannover.de}

\author[Hannover]{Insa Neuweiler}
\author[Barcelona]{Jesus Carrera}
\author[Barcelona]{Marco Dentz}

\address[Hannover]{Institute of Fluid Mechanics in Civil Engineering, Leibniz Universit\"{a}t Hannover, Hannover, Germany}
\address[Barcelona]{Institute of Environmental Assessment and Water Research (ID\AE A), Spanish Council of Scientific Research (CSIC), Barcelona, Spain}

\begin{abstract}
We study modeling of two-phase flow in highly heterogeneous fractured and porous media. The flow behaviour is strongly influenced by mass transfer between a highly permeable (mobile) fracture domain and less permeable (immobile) matrix blocks.
We quantify the effective two-phase flow behaviour using a multirate rate mass transfer (MRMT) approach. We discuss the range of
applicability of the MRMT approach in
terms of the pertinent viscous and capillary diffusion time scales.
We scrutinize the linearization of capillary
diffusion in the immobile regions, which allows for the formulation of
MRMT in the form of a non-local single equation model. The global
memory function, which encodes mass transfer between the mobile and
the immobile regions, is at the center of this method. We propose two
methods to estimate the global memory function for a fracture network with given fracture and
matrix geometry. Both employ a scaling approach based on
the known local memory function for a given immobile region. With the first method, the local memory function is calculated numerically, while the
second one employs a parametric memory function in form of truncated
power-law. The developed concepts are applied and tested for fracture
networks of different complexity. We find that both physically based
parameter estimation methods for the global memory function provide
predictive MRMT approaches for the description of multiphase flow in
highly heterogeneous porous media.
\end{abstract}


\end{frontmatter}


\section{Introduction}

Flow and transport through highly heterogeneous porous and fractured
media may be poorly predicted with equivalent
homogeneous models that are characterized by effective hydraulic
parameters. The impact of heterogeneity manifests in heavy
tails of solute breakthrough curves or recovery curves during
water flooding of an oil reservoir which is not captured well by such models. This behavior may originate from
the local non-equilibrium of the flow or transport processes which are
observed on a large scale.

Transport behavior caused by local non-equilibrium has been investigated for
solute \cite{Zinn2004,mckenna2001tracer,haggerty2004controls,kang2015impact}
and heat \cite{Geiger2010,klepikova2014passive} transport in groundwater
and for two phase flow processes like oil recovery \cite{Gilman1983}
or unsaturated flow \cite{neuweiler2012non}.
When flow and transport is considered in a medium where non-equilibrium behavior
is relevant, the full porous medium structure needs in principle to be
represented in a flow or transport model, in order to capture the local processes.
For a fractured rock this means that fractures and rock matrix need to be resolved in a model, which is computationally very expensive. As, however, tailing often needs to be captured in a model prediction, large effort has been spent to develop simplified modeling approaches that are less expensive, but can still capture the main features. This applies also to two-phase flow problems, such as modeling of oil recovery.

The classical approaches on Darcy scale to model flow and transport
in fracture networks are the equivalent porous media (EPM) approach,
the discrete fracture (DF) approach and the multi-continuum approach
\cite{bear1990introduction,bear1993flow}. For the EPM,
the fractures are modeled together with the surrounding rock matrix as
an equivalent porous media \cite{adler2012fractured}
and are thus represented by a variation of flow or transport parameters of
the rock matrix locally. These approaches are often too simplified to capture tailing effects.
For the DF approach, the rock matrix continuum is superimposed
by lower dimensional elements, which represent the fractures. When
applying a numerical scheme to solve the resulting models for the DF approach, a fine discretization is required to accurately
represent the local differences in flow and transport parameters,
especially at the interfaces between matrix and fractures \cite{ref:didonato_2004}.

Multi-continua approaches conceptualize the fractured medium in terms
of a mobile primary continuum, the fracture network, and a series of less
mobile secondary continua, the matrix blocks, which communicate with
the primary continuum through properly posed continuity conditions at
the interfaces between the continua. Mass transfer processes in the primary
continuum equilibrate fast over the scale of a representative
elementary volume and therefore are represented through a spatial
average. Thus, the complex spatial structure of the fracture network is
represented in an effective way as an equivalent porous medium, while
the slower processes in the secondary continua participate in the
average mass transfer in terms of non-local sink-source terms.
Common multi-continua models \cite{simuunek2008,szymkiewicz2012,ref:gerke_2006,bear1993flow} include
the dual continua (DC) and triple continua (TC) \cite{wu2006multiple}
and the multiple interacting continua (MINC) models \cite{pruess1992brief,ref:tatomir2011}.

Dual porosity (DP) models or mobile-immobile
models are a special case of the DC model. In fractured rock it is obvious
to distinguish two zones: the fracture network and the matrix blocks.
The fracture network represents the mobile zone, where ``fast''
flow or transport processes takes place, and the matrix blocks represent
immobile zones, characterized by slow exchange processes. When the
fracture network is sufficiently well connected \cite{pruess1992brief}, so that
an REV for the fracture network can be defined \cite{bear1990introduction}, it is feasible
to model the fracture network and the matrix blocks by applying a mass balance
for each zone separately. The crucial point in this type of
models is the modeling of the processes in the immobile zone respectively
the exchange fluxes between mobile and immobile zone. The DP model
can be solved numerically for each zone directly \cite{andersen2014model}
or can be simplified using transfer functions or the multirate mass
transfer (MRMT) approach.

In the transfer function approach (for example \cite{ref:Ramirez2009,ref:AlKobaisi2009,ref:Abushaikha2008}), the exchange
between mobile and immobile zone is approximated as a linear or nonlinear
single rate mass transfer and the immobile zone is represented as a fully mixed system. Different types of transfer functions have
been reported for diffusive solute transport, capillary counter-current flow during two-phase flow or flow driven
by gravity \cite{WarrenRoot,vermeulen1953theory,DiDonato2007}.

A MRMT approach is obtained if the response of the immobile zone is
modeled by a sink / source term that is non-local in time. The kernel
of the time integration can be considered a memory
function. The memory function can be expanded into
a sum of exponential functions \cite{arbogast1992simplified,Willmann2010}.
Exponential functions are the analytical solution for single rate transfer
processes. This means that, for example, diffusion like mass transfer
can be modeled by a distribution of single rate transfer
processes, hence MRMT model \cite{RoyHaggerty1995,Carrera1998}.

The MRMT and transfer function approaches differ from each other when applied to fracture networks.
For the transfer function approach there is one transfer function or one
single rate transfer process for each block size \cite{DiDonato2007}.
The block size distribution is then modeled by a superposition of
transfer functions and the resulting model is referred as multi rate
double porosity (MRDP) model. MRMT models can also be represented by superpositions, however, the
superposition does not necessarily represent a superposition over
block sizes, but the superposition is also due to the expansion of the
memory function into exponentials to capture dynamics different from single rate transfer.
Naturally, a superposition over different block sizes can be carried out on top of that.

Despite the variety of different approaches to model two-phase flow in fractures in a simplified way, while capturing the effect of the heterogeneous structure, they are not much applied in practice. One reason is that more parameters are required than for a two-phase flow model in a homogeneous medium and it is not so clear how these parameters could be estimated. This is in particular true if fracture networks are considered. Also, concepts are often tested with media with one single fracture (for example \cite{tecklenburg2013,andersen2014model}), but not often for fracture networks. Geiger et al. \cite{Geiger2011} studied a fracture network without taking heterogeneity in the capillary forces into account.

In this contribution we show the applicability of a MRMT model (presented in \cite{tecklenburg2013}) for
immiscible two phase flow to two dimensional fracture networks, where the flow in the fracture network is simplified by a single continuum. We present timescales for characterizing flow in fracture networks to quantify conditions where the MRMT model is needed to make good predictions of recovery. We also make suggestions how parameters of the MRMT model for a fracture network can be estimated. The parameters are calculated by analyzing fracture and matrix geometry. We show two approaches
to approximate the global memory functions for the MRMT model, both of them based on a superposition of functions obtained from a scaling of a reference memory function. In the one case the reference memory function is calculated numerically and in the other case it is approximated by a truncated power law function.
To demonstrate the methods, we consider a forced imbibition scenario in a fracture
network and spontaneous imbibition driven by capillary pressure in
the matrix blocks. We compare the results
of the one dimensional MRMT model with estimated parameters and the results of a full two dimensional
two phase flow model applying the EPM approach, where the code Dumux \cite{Flemisch20111102} is used.

Although we focus on fractured media, all concepts can be easily transferred to highly heterogeneous porous media in general. The heterogeneous medium would be split into a 'fast' domain (equivalent to the fracture domain) and a 'stagnant' domain (equivalent to the matrix domain).

This paper is organized as follows. In the second section we show
the equations for two phase flow in porous media and the upscaled MRMT model. In section 3, we discuss a scaling approach to calculate the parameters for the MRMT model.
In section 4, we apply the MRMT model with estimated parameters to different fracture networks and conclude
with a discussion. In the Appendix, we show how to numerically solve the MRMT as MRDP model.

\begin{figure}
\noindent \begin{centering}
\includegraphics[width=10cm]{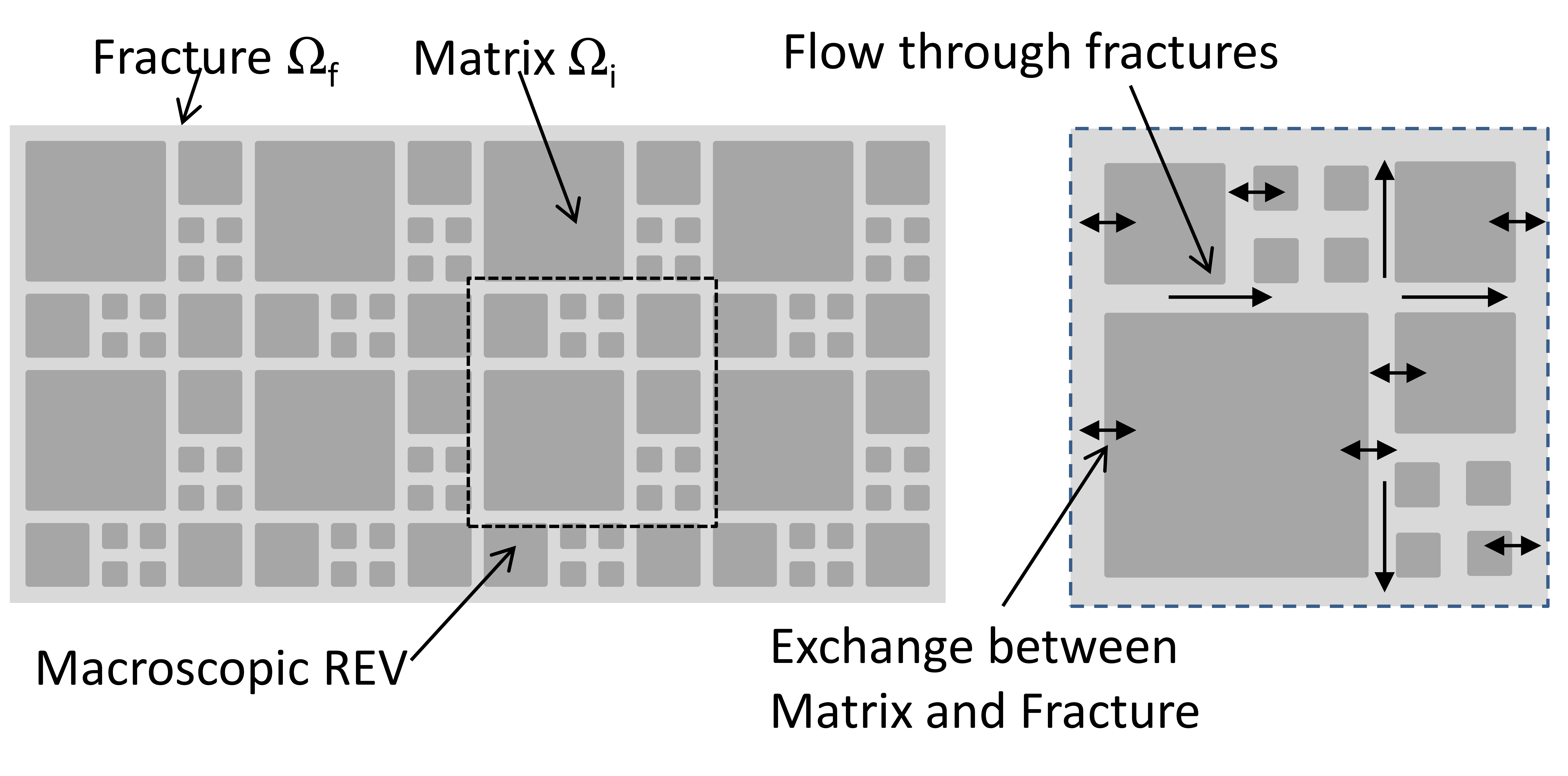}
\par\end{centering}

\protect\caption{\label{fig:schematic-reprasentation}schematic representation of fractured
media.}
\end{figure}

\section{Two-Phase Flow Multirate Mass Transfer Approach}

In this section, we outline the two-phase flow equations that describe
immiscible displacement in fractured and porous media. We then
generalize the multi-rate mass transfer model derived in
\cite{tecklenburg2013} to more realistic media, which are
characterized by a distribution of immobile regions with different
characteristic capillary diffusion time scales. 
\subsection{Two-Phase Flow in Porous Media}
We model the horizontal flow of two incompressible and immiscible fluids in a
rigid porous media. Based on the surface tension, one fluid
is referred to as the wetting phase (index $w$, e.g.,  water) and the
other one as the non-wetting phase (index $nw$, e.g., oil). Each phase
is described by a phase pressure $p_\alpha$ ($\textrm{M\ensuremath{L^{-1}T^{-2}}}$)
and normalized saturation $S_\alpha$ (-), with $\alpha = w,nw$, which fulfill
$S_{nw} + S_w = 1$ .

Conservation of mass for each fluid is expressed as
\begin{subequations}
\label{full:model}
\begin{align}
n_{f}\frac{\partial\left(\rho_{\alpha}S_{\alpha}\right)}{\partial t} +
\nabla\cdot\left(\rho_{\alpha}\mathbf{q_{\alpha}}\right)=0,
\end{align}
where $n_{f}$ (-) is porosity.  The specific discharge $\mathbf{q}_\alpha$
($\textrm{L\ensuremath{T^{-1}}}$) for each
phase is modeled by Darcy's law neglecting gravity
\begin{align}
\mathbf{q}_{\alpha}=-\frac{\mathbf{K} k_{r,\alpha}}{\mu_{\alpha}}\nabla
p_{\alpha} \label{eq:fullmodel}
\end{align}
where $\mathbf{K}$ ($\textrm{\ensuremath{L^{2}}}$)
is the intrinsic permeability tensor, $k_{r,\alpha}$ (-) is the relative
permeability of phase $\alpha$, and
$\mu_\alpha$ its viscosity ($\textrm{M\ensuremath{L^{-1}T^{-1}}}$).
The capillary pressure is defined as the pressure difference between
non-wetting and wetting phase and is assumed to be a unique function
of saturation,
\begin{align}
p_{c}(S_{w})=p_{nw}-p_{w}.\label{eq:pc}
\end{align}
\end{subequations}
Equations~\eqref{full:model} are referred below
as the full model. They are the starting point for the derivation
of the MRMT model, which is outlined below. A detailed derivation
from the full model to the MRMT model by homogenization theory is
presented in \cite{tecklenburg2013}.

By introducing the total Darcy velocity
$\mathbf{q}_{t}=\mathbf{q}_{nw}+\mathbf{q}_{w}$,
the full model~\eqref{full:model} can be cast in the fractional
flow formulation for the saturation of the wetting phase $S \equiv
S_w$ as in~\cite{Bear:1972}
\begin{align}
n_{f}\frac{\partial S}{\partial t} +
\nabla\cdot\left(\mathbf{q}_{t} f \right) -
\nabla\cdot\left(\frac{\mathbf{K}}{\mu_{nw}} \Lambda\nabla
  p_{c}\right)=0, \label{eq:fracflowsat}
\end{align}
where the fractional flow functions are defined as $f=\lambda_{w}/\lambda_{t}$
and $\Lambda = k_{r,nw}f$. The phase mobilities are $\lambda_{\alpha} = k_{r,\alpha}/ \mu_{\alpha}$
and the total mobility is $\lambda_{t} = \lambda_{w}+\lambda_{nw}$.
The incompressibility of each fluid phase gives for the total Darcy velocity $\nabla\cdot\mathbf{q}_{t}=0$.

\subsection{Multirate Mass Transfer (MRMT) Model}
As outlined in the Introduction, the MRMT model divides the fractured
media into a mobile zone $\Omega_f$, the connected primary continuum, and a suite
of immobile zones $\Omega_i$, the secondary continua. Quantities
referring to the fracture domain in the following are marked by the
subscript $f$, while quantities referring 
to the matrix domain are denoted by a subscript $m$. Quantities
referring to the secondary continuum $\omega$ are marked by the subscript
$\omega$.

In the fracture network the flow is assumed to be viscous dominated,
while in the matrix blocks spontaneous countercurrent
imbibition is assumed to be the dominating flow process, so that
viscous flow can be neglected. The fracture and matrix continua
communicate through the continuity of
capillary pressure and fluxes at their interface.

The non-local two phase flow model presented in~\cite{tecklenburg2013}
is derived from~\eqref{eq:fracflowsat}
by applying homogenization theory to a media which consists of a
fracture continuum and a single class of immobile continua
characterized by the same geometry and physical characteristics.
We generalize here this non-local model to account for mass exchange
between a mobile continuum and a spectrum of immobile continua
characterized by multiple exchange rates. The mass balance equation for
the average fracture saturation $S_f$ in such a medium is given by
\begin{align}
& \frac{\partial S_f}{\partial t}  +
\frac{{\mathbf{q}}_{t,f}}{n_{ff}}  \cdot
\mathbf{\nabla} f_f
=
- a_v \int d \omega \mathcal P(\omega) \chi_\omega \frac{\partial}{\partial t} \langle S_\omega \rangle,
\label{eq:buckley}
\end{align}
where the angular brackets denote the average over the immobile zone
of type $\omega$ (cf. the single squares in figure 1),
\begin{align}
\langle S_\omega \rangle = \frac{1}{V_{\omega}} \int\limits_{\Omega_{\omega}} d \mathbf x
S_{\omega}.
\end{align}
The distribution density $\mathcal P(\omega)$ measures the relative
frequency of immobile zones of type $\omega$ which are characterized
by the same geometry and physical properties. The average total Darcy velocity in the fracture continuum is denoted by
$\mathbf q_{t,f}$.
We defined the effective volume ratio $a_v = \frac{V_m n_{f,m}}{V_f n_{f,f}}$ with $n_{f,f}$
intrinsic porosity and $V_f$ total volume of the fracture domain, and
$n_{f,m}$ and $V_m$ the average porosity and total volume of the
matrix domain. The effective volume ratio of the immobile zone of type
$\omega$ is given by $\chi_\omega = \frac{V_\omega n_{f,\omega}}{ V_m
  n_{f,m}}$ with $n_{f,\omega}$ the intrinsic porosity of the immobile
zone of type $\omega$.

The saturation $S_\omega$ of the immobile continuum $\omega$ is given
by the mass conservation equation
\begin{align}
n_{f,m} \frac{\partial S_\omega}{\partial t}
- \mathbf \nabla \cdot[ \mathcal D_\omega^\prime \mathbf \nabla S_\omega] = 0,
\label{eq:matrixdiffusion}
\end{align}
where the non-linear capillary
diffusivity is defined as in~\cite[][]{ref:mcworther_1990}
\begin{align}
\label{Dnl}
\mathcal D_\omega^\prime = - \frac{K_\omega}{\mu_{nw}} \Lambda_\omega \frac{dp_{c,\omega}}{dS_{\omega}}.
\end{align}
The capillary pressure-saturation relationship in the immobile zones
of type $\omega$ is given by $p_{c,\omega}$, which is proportional to
$p_{c,\omega} \sim 1/\sqrt{K_\omega}$ (Leverett scaling) with
$K_\omega$ the permeability of immobiles zones of type
$\omega$.  As a result, we have the scaling of $\mathcal
D_\omega^\prime \sim K_\omega^{1/2}$ with permeability.

The boundary conditions for each matrix block $i$ are imposed by the saturation in the surrounding fracture
\begin{align}
p_{c,f}(S_f) = p_{c,\omega}(S_\omega), && \mathbf x \in \partial \Omega_\omega,
\label{bceffmulti}
\end{align}
i.e., continuous capillary pressure over the interface
between fracture and matrix, which thus assumes local capillary equilibrium
in the medium.

The initial saturation is denoted by $S_{\omega}(t=0)=S_{\omega}^{0}$.
Equation (\ref{eq:matrixdiffusion}) quantifies the spontaneous imbibition of the wetting fluid
into the matrix and needs to be solved for the given geometry of the individual matrix
block.
It is a non-linear diffusion equation, whose solutions have the
characteristic Boltzmann scaling, i.e., a saturation front scales with
$\sqrt{t}$. As outlined in~\cite{tecklenburg2013}, the flux term
in~\eqref{eq:matrixdiffusion} is linearized by approximating the capillary diffusivity $\mathcal
D_\omega^\prime$ by a suitably chosen constant value, $\mathcal D_\omega^\prime
\approx \mathcal D_\omega =$ constant. The determination of an
equivalent constant diffusion coefficient from the non-linear
diffusion problem is discussed in
Section~\ref{sec:capillarytimescales}. This substitution yields a linear diffusion
equation for the saturation $S_\omega$ of immobile zones of type $\omega$,
\begin{align}
n_{f\omega} \frac{\partial S_\omega}{\partial t} -
\mathbf{\nabla} \cdot [\mathcal D_\omega \mathbf \nabla  S_\omega] = 0.
\label{eqim3}
\end{align}
The linearized diffusion equation~\eqref{eqim3} can be solved using
the method of Green's functions~\cite[][]{tecklenburg2013}. This gives
for the spatially averaged saturation in regions $\omega$
\begin{align}
\langle S_\omega \rangle = \langle S_\omega^0 \rangle -
\int\limits_0^t dt^\prime \left\langle S_m^0(\mathbf x) g_{\omega}(\mathbf
x,t^\prime) \right\rangle + \int\limits_{0}^t
d t^\prime \varphi_{\omega}(t - t^\prime) S_{\omega,b}(\mathbf x,t^\prime)
,
\label{Smav}
\end{align}
where the boundary saturation $S_{\omega,b}(\mathbf x,t) =
P_{c,\omega}^{-1}\{P_{c,f}[S_f(\mathbf x,t)]\}$ is enforced by continuity of capillary pressure.
The Green's function $g_\omega(\mathbf x,t)$
solves the linearized problem~\eqref{eqim3} for the unit pulse $g_\omega(\mathbf x,t)|_{\mathbf x
\in \partial \Omega_\omega} = \delta(t)$ on the boundary and the initial
condition of $g_\omega(\mathbf x, t = 0) = 0$. The local memory function is defined by the spatial average over the
Green's function as
\begin{align}
\label{phi}
\varphi_\omega(t) = \langle g_\omega(\mathbf x,t) \rangle.
\end{align}
Inserting~\eqref{Smav} into the right side of~\eqref{eq:buckley} gives
the following closed form equation for $S_f$
\begin{align}
\label{upscaled:model}
\frac{\partial S_f }{\partial t} + \frac{\mathbf{q}_{tf}}{n_{ff}} \cdot
\mathbf{\nabla} f_f
= - a_v \frac{\partial}{\partial t} \int\limits_{0}^t
d t^\prime \varphi(t - t^\prime)
S_{b}(\mathbf x,t^\prime)
+ a_v \int d \omega \mathcal P(\omega) \chi_\omega \langle  S_\omega^0 g \rangle,
\end{align}
where we defined the global memory function $\varphi(t)$ by
\begin{align}
\label{phiglobal}
\varphi(t) = \int d \omega \mathcal P(\omega) \chi_\omega \varphi_\omega(t).
\end{align}
Explicit analytical solutions for the local memory functions $\varphi_\omega(t)$
can be obtained for slabs, cylinders and
spheres~\cite[][]{RoyHaggerty1995,Carrera1998}.
For more complex geometries of the matrix domain $\Omega_\omega$, the local
memory functions can be obtained numerically~\cite[][]{Noetinger:2000, gouze08b}.
Eq. (\ref{upscaled:model}) is the upscaled model for the full problem described in Section 2.1. It has only one primary variable, the fracture saturation $S_f$, and it is parameterized with the two-phase flow parameters for the fracture and with the memory function $\varphi(t)$. The memory function is thus the core of the upscaled model.
\section{The Global Memory Function}

In the following we discuss the characteristic time scales of the
mobile fracture domain and the immobile matrix domains, first in the
light of the application limits of the MRMT approach, and second to
relate the geometrical and physical properties of the immobile domains
to the global memory function.

\subsection{Time Scales}
In this subsection, we discuss the conditions for the assumptions made in
the MRMT model posed in the previous section: Viscous dominated
fast flow in the fracture network and capillary dominated slow flow
in the matrix blocks. To this end, we suggest a definition of time
scales, which characterize fracture and matrix flow. 

\subsubsection{Viscous Time Scales}

The characteristic time scale for the viscosity dominated fluid
displacement in the fracture domain is given by
\begin{equation}
t_{q,f}=\frac{V_{f}n_{f,f} \langle S_f \rangle}{A_{f}q_{t,b}},\label{eq:tqf}
\end{equation}
where $A_{f}$ is the cross-sectional area of
the fracture network available for injection, and $q_{t,b}$ is
the total flux at the inlet boundary, and $\langle S_{f} \rangle$ the saturation
at breakthrough averaged over the whole fracture domain.
The viscous time scale $t_{q,f}$ compares the volume available for the
wetting fluid to the the volumetric flow rate in the fracture
domain. This corresponds to the characteristic breakthrough time at the
outlet.

The mean saturation $\langle S_{f} \rangle$ before breakthrough can be
found by analyzing the fractional flow function as described in
\cite{donaldson1985enhanced} or can be estimated by
$\langle {S}_{f} \rangle =1$, when the rarefaction wave of the displacement
solution is negligible.
The viscous time scale $t_{q,m}$ for the matrix blocks may be related to
$t_{q,f}$ by $t_{q,m}=(K_{f}/K_{m})  t_{q,f}$. Thus, viscous flow in the matrix
is negligible for $(K_f/K_m) \gg 1$, provided that flow in the fractures is viscous dominated.

\subsubsection{Capillary Diffusion Time Scales}
\label{sec:capillarytimescales}

The capillary diffusion time scales $t_{\omega}$ of an immobile region
of type $\omega$ can be
estimated from an analytical solution for countercurrent imbibition
as described by the non-linear diffusion
equation~\eqref{eq:matrixdiffusion} with the boundary conditions
given by~\eqref{bceffmulti}. As pointed out above, a saturation front displays the characteristic
$\sqrt{t}$ scaling of a diffusive front. McWorther and Sunada \cite{ref:mcworther_1990}
give an analytical solution for the displacement of a saturation isoline
for imbibition into a slab characterized by porosity $n_{f,\omega}$ and
diffusivity $\mathcal D^\prime_\omega$,
\begin{equation}
\label{xs}
r_\omega(S_\omega,t) = \frac{2 A_\omega}{n_{f,\omega}}\frac{\partial
F_\omega}{\partial S_\omega}\sqrt{t},
\end{equation}
where $r$ denotes the position of the invading front. It is specific
for a given saturation. The functions $F$ and $A$ are defined in \cite{ref:mcworther_1990}
implicitly by
\begin{align}
F_\omega(S_\omega) =1-\frac{\displaystyle\int\limits_{S_\omega}^{S_{b,\omega}} (\beta - S) \frac{\mathcal
    D_\omega^\prime(S)}{F_\omega(S)}\mathrm{d}\beta}{\displaystyle\int\limits_{S_{0,\omega}}^{S_{b,\omega}}(\beta-S_{0,\omega})
  \frac{\mathcal D_\omega^\prime(\beta)}{F_\omega(\beta)}\mathrm{d}\beta}, &&
A_\omega^{2} =
\frac{n_{f,\omega}}{2}\int\limits_{S_{0,\omega}}^{S_{b,\omega}}(\beta-S_{0,\omega})
\frac{\mathcal D_\omega^\prime(\beta)}{F_\omega(\beta)} \mathrm{d}\beta
\end{align}
We obtain a characteristic front depth $\overline{r}_\omega$, and with that a characteristic
time scale by averaging~\eqref{xs} over saturation from the
initial saturation $S_0$ to the boundary saturation $S_b$, which gives
\begin{align}
\label{Ddef}
\overline r_\omega(t) = \sqrt{\frac{2 \mathcal D_\omega t}{n_{f,\omega}}}, && \mathcal D_\omega =
\frac{n_{f,\omega}}{2}
\left(\frac{2A_\omega}{n_{f,\omega}} \frac{F_{b,\omega} - F_{0,\omega}}{S_{b,\omega} -
S_{0,\omega}} \right)^2,
\end{align}
where we set $F_{b,\omega} = F_\omega(S_{b,\omega})$ and $F_{0,\omega} = F(S_{0,\omega})$. Equation~\eqref{Ddef}
sets the equivalent constant diffusivity $\mathcal D_\omega$
in~\eqref{eqim3}. Note that $\mathcal D_\omega$ scales with hydraulic
conductivity as $\sim K_\omega^{1/2}$.

We define now the characteristic capillary diffusion time scale as time $t_\omega$ at
which the average front has reached the length scale $L_\omega$ such that
\begin{equation}
t_{\omega} = \frac{n_{f,\omega} L_\omega^2}{2 \mathcal D_\omega}
\end{equation}
%
Notice that this capillary diffusion time scale is derived from a 1d solution, where counter current flow into one block happens from two sides. In a 2d block, where counter current flow occurs from four sides,
the capillary diffusion time scale yields
\begin{equation}
t_{\omega} = \frac{n_{f,\omega} L_\omega^2}{4 \mathcal D_\omega}\label{eq:td_thiswork}
\end{equation}
The characteristic length scale for the matrix is the shortest
distance that is passed by the capillary counter
current displacement front, until two front positions meet. It is
can be estimated by \cite{ma1997}
\begin{equation}
L_{\omega}=\sqrt{\frac{V_\omega}{\sum_{i=1}^{N} \mathcal
    A_{i}/l_{\mathcal A_{i}}}},
\label{eq:Ma}
\end{equation}
where $V_\omega$ is the volume of the immobile region of type
$\omega$. It has
$N$ faces $i = 1,\dots,N$, each of which has the area $\mathcal A_{i}$ and a
distance $l_{\mathcal A_{i}}$ to the volumetric center of the
matrix block. For a two dimensional rectangular block with a width $a$
and length $b$ one obtains
\begin{equation} 
L_{\omega}=\sqrt{\frac{1}{4}\frac{a^{2}b^{2}}{a^{2}+b^{2}}}. \label{eq:Mablock}
\end{equation}
Note that this implies that matrix blocks can be anisotropic.

For the MRMT model, the non-equilibrium aspects of the immobile zone,
this means the time behavior before it equilibrates with the mobile
zone, is important because this is when the matrix-fracture transfer
strongly influences the front speed in the fracture network. We
therefore use the length scale~\eqref{eq:Ma} to derive the capillary
diffusion time scale $t_{\omega}$ ~\eqref{eq:td_thiswork}, which reflects this
non-equilibrium aspects well. This is discussed in more detail in
 \ref{comparison:timescales}. 


\subsection{The Memory Function \label{sub:Calculating-the-memory}}

The global memory function $\varphi(t)$ in \eqref{phiglobal} is at the heart of the MRMT modeling
approach. It encodes the mass transfer between mobile and immobile
zones, and its structure reflects the geometry and heterogeneity of the
immobile regions~\cite[][]{gouze08b} that are no longer resolved in the upscaled model.
As indicated by~\eqref{phi} and~\eqref{phiglobal}, the global memory
function can in principle be determined by (i) calculating the local memory
functions $\varphi_{\omega}(t)$ for the matrix geometry of each block type $\omega$ of the fracture
network, and (ii) determining the weighted sum over the local memory
functions using the volume fractions of each type $\omega$ of immobile
zones.

The most general and accurate method would be to (numerically)
determine the local memory function $\varphi_{\omega}(t)$ for each matrix geometry and to average them.
Although the calculation of the local memory functions needs to be done only
once at the beginning of the simulation, this method is not very practical.
Below, we present two different ways to estimate the local memory functions. Both methods are based on a scaling approach for the local memory functions $\varphi_{\omega}(t)$. This reduces the effort to calculating only one local reference memory function $\hat{\varphi}(t)$. The distribution over the local functions is obtained by rescaling the reference function with the capillary diffusion time scales $t_{\omega}$ obtained from the matrix geometry of the single blocks. The local reference memory function $\hat{\varphi}(t)$ is here calculated with two different methods.
For the first method, it is calculated on a reference block numerically. The second
method is based on approximating the local memory functions by an analytical
function.


The local memory functions $\varphi_{\omega}(t)$ scale
with respect to the characteristic capillary diffusion time scale $t_{\omega}$
determined in the previous section. This can be seen as follows.
Equation~\eqref{eqim3} for the Green function $g_\omega(\mathbf x,t)$
can be non-dimensionalized by scaling $\mathbf x = \mathbf x^\prime
L_\omega / \sqrt{2}$ and $t = t_\omega t^\prime$ with $L_\omega$ and $t_\omega$
given by~\eqref{eq:Ma} and~\eqref{eq:td_thiswork}.
Thus we obtain for the Green function the governing equation
\begin{align}
\frac{\partial g^\prime_\omega}{\partial t} - \mathbf
{\nabla^\prime}^2  g^\prime_\omega = 0, && g_\omega(\mathbf x,t) =
t_\omega^{-1} g^\prime_\omega(\mathbf x \sqrt{2}/L_\omega,t/t_\omega),
\label{eqim3lin}
\end{align}
with the boundary condition $g^\prime(\mathbf
x^\prime,t^\prime)|_{\mathbf x \in \partial \Omega^\prime} =
\delta(t^\prime)$.
The local memory function is then given by
\begin{align}
\label{phiscaling}
\varphi_\omega(t) = \frac{1}{t_\omega} \left[ \frac{1}{2^{d/2}V_\omega^\prime} \int\limits_{\Omega^\prime_\omega} d
\mathbf x^\prime  g^\prime_\omega(\mathbf x^\prime,t/t_\omega) \right] \approx
\frac{1}{t_\omega} \hat \varphi(t/t_\omega).
\end{align}
Note that the term in the square brackets depends on the properties of
the immobile zone predominantly through the time scale
$t_\omega$. If the immobile zones have the same topological
characteristics, which we assume is the case, we can in good approximation disregard the specificities of the
particular geometry of the immobile zones. Thus we approximate the
memory function by the scaling form on the right side
of~\eqref{phiscaling}, which defines the scaling function $\hat
\varphi(t)$. This scaling function can then be estimated from the
memory function for a single immobile domain $\omega_0$ as
\begin{align}
\label{scaling}
\hat \varphi(t') = t_{\omega_0} \varphi_{\omega_0}(t_{\omega_0} t').
\end{align}
The memory function for an immobile domain $\omega_0$ behaves as
$1/\sqrt{t}$ for $t \ll t_{\omega_0}$ and decreases exponentially fast for
$t \gg t_{\omega_0}$~\cite[][]{RoyHaggerty1995, Carrera1998}. Thus, the
scaling function $\hat \varphi(t')$ shows the $1/\sqrt{t'}$ decay for $t'
\ll 1$ and a sharper decrease for $t' \gg 1$.

The global memory function is now given in terms of the scaling function
$\hat \varphi(t')$ as
\begin{align}
\label{eq:scaling}
\varphi(t) = \int d t_{\omega} \mathcal P_t(t_{\omega}) \frac{1}{t_{\omega}} \hat
\varphi(t/t_{\omega}).
\end{align}
where $\mathcal P_t(t_{\omega})$ denotes the probability density
function (PDF) of characteristic capillary time scales.
It can be expressed in terms of the distributions of the geometric and
physical characteristics of the immobile zones through having the
explicit expression~\eqref{eq:td_thiswork} for $t_\omega$. In practice, the integral will be transformed to a sum over all matrix block types.

If the matrix blocks have the same permeability $K_{\omega}$, the dependency of the local memory functions on the capillary diffusion time scale $t_{\omega}$ can be reformulated into a dependency on the characteristic length scale $L_{\omega}$, as defined in eq. (\ref{eq:Ma}). The local memory function $\varphi_{\omega}$ can then be obtained from the local memory function $\varphi_{\omega_0}$
\begin{equation}
\varphi_{\omega}(t)= \frac{1}{t_{\omega}} \hat\varphi(t/t_{\omega}) = \frac{t_{\omega_0}}{t_{\omega}} \varphi_{\omega_0} \left(t\frac{t_{\omega_0}}{t_{\omega}}\right) = \frac{L_{\omega_0}^{2}}{L_{\omega}^{2}} \varphi_{\omega_0} \left(t\frac{L_{\omega_0}^{2}}{L_{\omega}^{2}}\right).
\end{equation}

In general, the PDF of time scales, $\mathcal P_t(t_{\omega})$, can be related to the PDF of matrix block properties, which can be estimated directly from information about the fracture network.
For simplicity, we assume that the intrinsic immobile porosities are the same for all regions such that $n_{f,\omega} = n_{f,m}$. Furthermore
we assume that the effective capillary diffusion depends on the
physical properties of the medium only through its dependence on the
hydraulic conductivity such that $\mathcal D_\omega = A
\sqrt{K_\omega}$ with $A$ a constant. The time scale $t_\omega$ then
reads as
\begin{align}
t_\omega = \frac{n_{f,m} L_{\omega}^2}{2 A \sqrt{K_\omega}}.
\end{align}
We write the in general coupled PDF $p_{L,K}(L,K)$ of $L_\omega$ and $K_\omega$ as
\begin{align}
p_{L,K}(L,K) = p_{K|L}(K|L) p_{L}(L),
\end{align}
where $p_{K|L}(K|L)$ is the conditional PDF of $K$ given $L$. Thus,
the PDF of time scales can be written as
\begin{align}
\mathcal P_t(t_{\omega}) = \sqrt{\frac{2 A}{n_{f,m} t_{\omega}}}
\int\limits_0^\infty d K K^{1/4} p_{K|L}(K|L) p_L\left(\sqrt{\frac{2 A
      \sqrt{K} t_{\omega}}{n_{f,m}}}\right).
\end{align}
If the immobile regions are all of equal conductivity but variable
extensions, the distribution of characteristic time scales simplifies
to
\begin{align}
\mathcal P_t(t_{\omega}) = \sqrt{\frac{2 A \sqrt{K}}{n_{f,m} t_{\omega}}}
p_L\left(\sqrt{\frac{2 A \sqrt{K} t_{\omega}}{n_{f,m}}}\right).
\end{align}

Note that the flow behavior is controlled by the distribution of mass
transfer time scales $\mathcal P_t(t_{\omega})$, in particular for large
times. Long residence times can be caused equally by low hydraulic
conductivities and large sizes of the immobile regions. For the sake
of computational efficiency, the test cases studied in the following
consider a distribution of characteristic matrix length scales
while the hydraulic conductivity is kept constant.

Besides the PDF for capillary diffusion time scale distribution, the reference memory function $\hat \varphi(t')$ needs to be determined in order to calculate the global memory function from eq. (\ref{eq:scaling}). We suggest here two different methods to do that.


In the first method, the local memory function $\varphi_{\omega_0}$ is obtained from the numerical solution
of the capillary diffusion problem~\eqref{eq:matrixdiffusion} for a  single
immobile region of type $\omega_0$. The pragmatical aspects are outlined in~\ref{num:memoryfunction}. We use then the scaling relation~\eqref{scaling}
in order to determine $\hat \varphi(t')$ from the numerically calculated local memory function
$\varphi_{\omega_0}$.

\begin{figure}
\begin{centering}
\includegraphics[width=10cm]{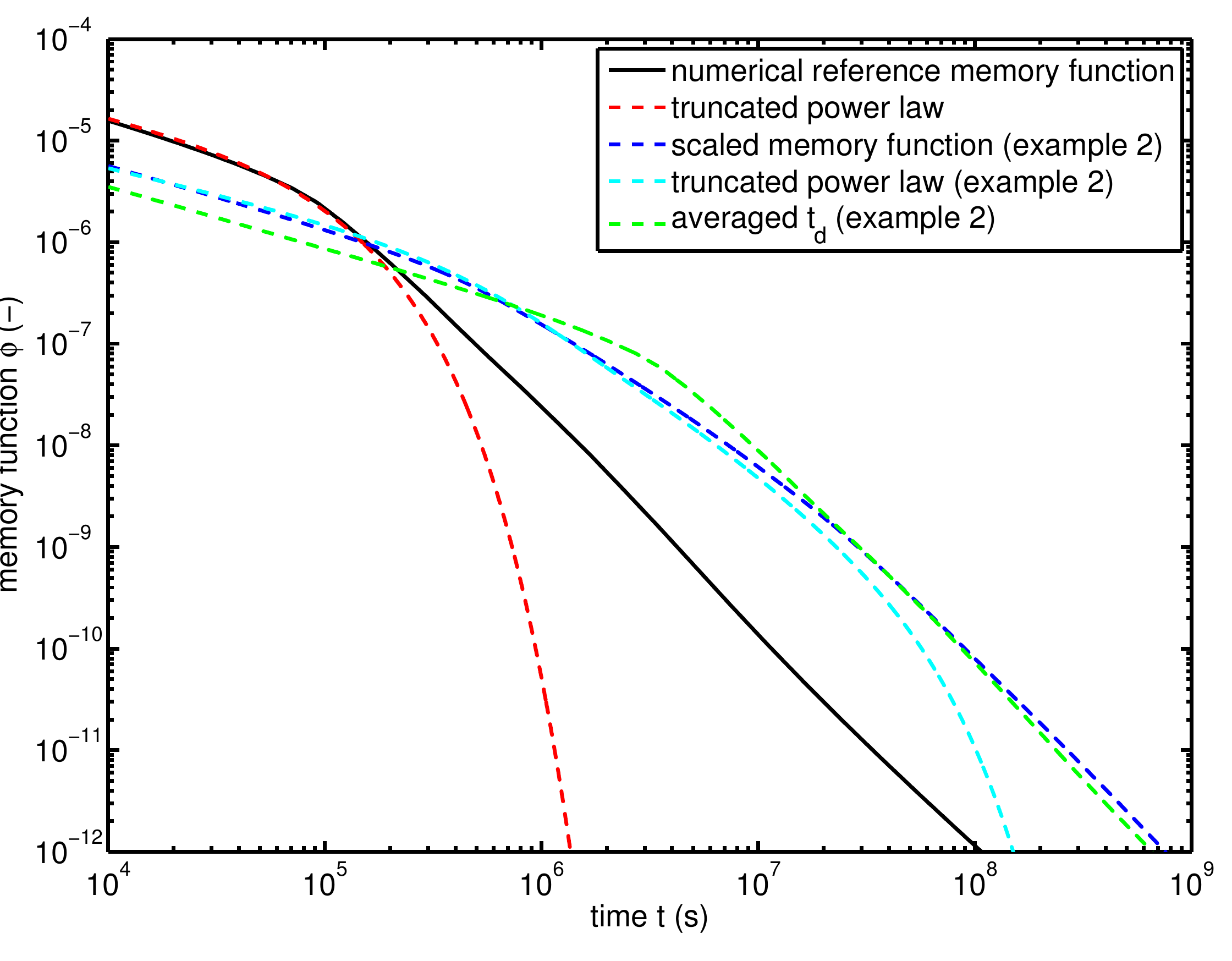}
\end{centering}

\caption{\label{fig:memoryfunctions}Memory functions with a single capillary diffusion time scale $t_{\omega}$ and memory functions for a distribution of capillary diffusion timescales from example  2 in section~\ref{sec:Test-cases} }
\end{figure}

The second method approximates $\hat{\varphi}(t')$ directly by the parametric form
\begin{align}
\label{eq:truncated_power_law}
\hat \varphi(t') =
\frac{\exp(-t')}{\Gamma(1/2)\sqrt{t'}},
\end{align}
where $\Gamma(x)$ denotes the Gamma function. The scaling
function~\eqref{eq:truncated_power_law} has similar characteristics as the scaling
function obtained from
the first method through~\eqref{scaling}, but a simple analytical
form. With this approximation, no numerical solutions are needed. Notice that the cut-off behavior of the local memory function
for the non-linear diffusion problem is different from the exponential
decay of the surrogate scaling form~\eqref{eq:truncated_power_law}. This is due to
the fact that the non-linear diffusion coefficient~\eqref{Dnl} decreases
with decreasing saturation. This leads to a smoother break-off
behavior than the exponential cut-off
of~\eqref{eq:truncated_power_law}~\cite[][]{tecklenburg2013}. This is
illustrated in figure~\ref{fig:memoryfunctions}, where a local memory function is shown together with its approximation by a truncated power law. Also shown in the figure are the global memory functions for a test example (discussed later), resulting from the superposition of the local ones.

%
%

\subsection{Flow regimes}

The MRMT model is strictly valid only for the case that flow in the
fractures is viscous dominated and that flow in the matrix is
capillary dominated, with a time scale that is comparable or larger
than the viscous time scale in the fractures.
To quantify flow regimes, the time scales defined by (\ref{eq:tqf})
for viscous forces and by (\ref{eq:td_thiswork}) for capillary forces have to be compared.

When the viscous time scale in the fracture network is much smaller
than the capillary diffusion time scale in the fracture network, i.e.,
$t_{q,f} \ll t_{d,f}$, viscous flow dominates and capillary forces
can be disregarded in the fractures. Whether viscous forces in the matrix
are relevant or not, depends on the permeability contrast, as
outlined above.

When the flow in the fracture is faster than countercurrent flow in
the matrix, $t_{q,f} < t_{\omega}$, the matrix
blocks are completely surrounded by the wetting phase in a short time
compared to the typical relaxation time for the matrix counter current
flow. If the time scales are in the same range, this condition is not fulfilled and the assumptions made to calculate the exchange fluxes are not exactly met. If the time scales are strongly separated ($t_{q,f} \ll t_{\omega}$) and one is only interested in the process for times smaller than $t_{\omega}$, one could avoid a MRMT model and simply model flow in the fracture network only.

In case that the flow in the matrix is faster than the flow
in the fracture, $t_{q,f} > t_{\omega}$, local equilibrium conditions hold and a single
equivalent porous medium describes the averaged flow well.
In this case a MRMT approach is not necessary.

\section{Study Cases\label{sec:Test-cases}}
We demonstrate the applicability of the MRMT model for two-phase flow
in fracture networks with two illustrative examples. With the the
first example we want to demonstrate the applicability of the MRMT model~\eqref{upscaled:model}
to fracture networks in general, using the criteria discussed above. With the second example,
we want to show that the model works for complex fracture networks
and that using an approximate memory function by scaling
a reference function either obtained numerically or approximated by a truncated
power law leads to useful predictions of recovery curves. To validate
the simulation results of the MRMT model, we
compare these results to the results of a full implicit 2D model~\eqref{full:model}
implemented in Dumux \cite{Flemisch20111102}.

\subsection{General Setup}

\begin{figure}
\includegraphics[width=10cm]{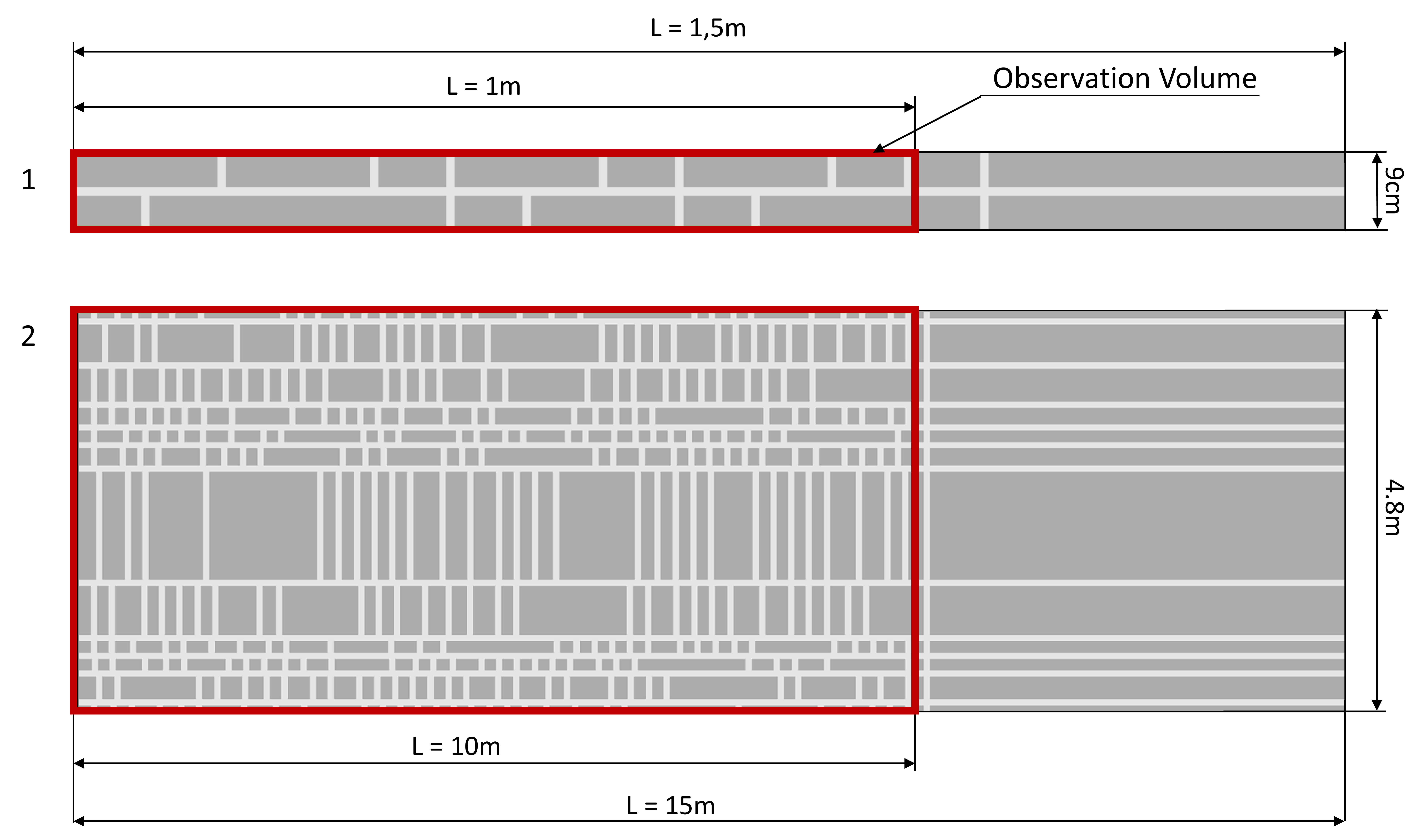}

\protect\caption{\label{fig:fracture-networks}Top: Domain 1 is
a simple fracture network for the first example. Bottom: Domain
2 is the complex fracture network for the second example. The fractures
are light gray and the matrix blocks are dark gray. We compare the
results of the models in the observation volume only to minimize boundary
effects. }
\end{figure}
 %

\begin{table}
\noindent \begin{centering}
\begin{tabular}{lllll}
\hline
 & \multicolumn{2}{c}{Example 1} & \multicolumn{2}{c}{Example 2}\tabularnewline
\hline
Parameter & Fracture & Matrix & Fracture & Matrix\tabularnewline
\hline
Intrinsic permeability $K_m$ (m$^2$) & $10^{-10}$ & see table
\ref{tab:Parameters-for-example-1} & $10^{-10}$ &
$10^{-14}$\tabularnewline
Porosity $n_{f}$ (-) & 0.1 & 0.1 & 0.1 & 0.1\tabularnewline
Residual non-wetting saturation $S_{nw,r}$ (-) & 0 & 0 & 0 & 0\tabularnewline
Residual wetting saturation $S_{w,r}$ (-) & 0 & 0 & 0 & 0\tabularnewline
van Genuchten parameter $1/\alpha$ (Pa) & 120 & 660 & 120 & 660\tabularnewline
van Genuchten parameter $N$ (-) & 1.75 & 1.75 & 1.75 & 1.75\tabularnewline
 &  &  &  & \tabularnewline
\hline
\end{tabular}
\par\end{centering}

\protect\caption{Spatial parameters for both examples. The capillary pressure curves
and the relative permeability are parameterized with the Mualem-Van
Genuchten model. \label{tab:Spatial-parameters}}
\end{table}

\begin{table}
\noindent \begin{centering}
\begin{tabular}{lllllll}
\hline
Setup &  & A \textbf{} & B\textbf{} & C1 & C2\textbf{}\tabularnewline
\hline
Intrinsic permeability $K_{m}$ $(\textrm{m}{}^{2})$ &  & $10^{-12}$ & $10^{-13}$ & $10^{-14}$ & $10^{-15}$\tabularnewline
viscous time scale $t_{q,f}$ (s) &  &  &  & $1.9\times10^{4}$ & \tabularnewline
capillary diffusion time scale $t_{d,f}$ (s) &  &  &  & $5.5\times10^{5}$ & \tabularnewline
capillary diffusion time scale $t_{\omega}$ (s) &  & $1.9\times10^{3}$ & $1.9\times10^{4}$ & $1.9\times10^{5}$ & $1.9\times10^{6}$ \tabularnewline
condition $t_{q,f}<t_{\omega}$ matched? &  & no & no & yes & yes \tabularnewline
\hline
\end{tabular}
\par\end{centering}

\protect\caption{\label{tab:Parameters-for-example-1}Intrinsic
  permeabilities of the different examples and the resulting process
  time scales. The typical length scales were chosen as the length of matrix blocks in orthogonal direction to the main flow direction.}
\end{table}

The fractured media for both examples are shown in figure \ref{fig:fracture-networks}
and the flow parameters can be found in table \ref{tab:Spatial-parameters}
and \ref{tab:Parameters-for-example-1}. For the examples we simulate
oil recovery from these fractured media during a water flooding of the
reservoir. Water (density $\rho=1000\,\mathrm{kg/m^{3}}$ , viscosity
$\mu=1.0\cdot10^{-3}\mathrm{Pa\cdot s}$) is the wetting fluid and oil ($\rho=890\,\mathrm{kg/m^{3}}$,
$\mu=8.0\cdot10^{-3}\mathrm{Pa\cdot s}$ ) is the non-wetting fluid.
At the beginning, the fractured media are almost filled with oil and
matrix and fracture are in capillary equilibrium. The water saturation
in the matrix blocks is $S_{m}^{0}=0.2$ and in the fracture network
$S_{f}^{0}=P_{c,f}^{-1}\{P_{c,m}(S_{m}^{0})\}=0.057$.

The boundary conditions for the fracture network are a constant inflow
for the wetting phase ($q_{w}=10^{-5}$m/s for the simple and $q_{w}=10^{-4}$m/s
for the complex fracture network) over the fracture sections of the left boundary and a no-flow boundary for the non-wetting
phase on the fracture at the left boundary. On the fracture at the
right boundary, there is a constant non-wetting pressure $p_{n}=10^{7}$Pa
and a constant water saturation of $S_{w}=0.2$. The upper and the
lower boundary and on all matrix boundaries are no-flow boundaries
for both phases. For the full 2D model the fracture networks as shown
in figure \ref{fig:fracture-networks} are used as parameter fields
using the boundary and initial conditions as described above. For the
MRMT model, we simulate an equivalent 1D fracture domain with homogeneous parameters.
The parameters for the MRMT model are calculated from the structure
in the observation volume (see figure \ref{fig:fracture-networks}).
An image analysis of the structure gives the total volumes $V_{f}$ and $V_{m}$
of the fracture respectively of the matrix, the cross-sectional area $A_{f}$
of the fracture network at the inlet as well as the width $a$ and length $b$
of each matrix block. These values are needed to calculate the parameters for the
MRMT model such as the effective volume ratio $a_{v}$, the total Darcy velocity $q_{t}$
and the local memory functions $\varphi_{\omega}$. The global memory function
is then calculated from all local memory functions using~\eqref{phiglobal} as
\begin{equation}\label{eq:globalmemory2}
\varphi(t)=\sum_{\omega} n_{\omega} \chi_{\omega}\varphi_{\omega}(t),
\end{equation}
where the integral in eq. (\ref{eq:scaling}) is now a sum and the
distribution function $\mathcal P$ is now the relative number of
blocks of type $\omega$, $n_{\omega}$. The boundary condition for the
MRMT model is a saturation of $S_{f}=1$ at the
left boundary. This boundary condition can be found from the fractional
flow function for a constant inflow for the wetting phase and a no-flow
boundary for the non-wetting phase.

For the MRMT model of the simple fracture network the fracture is discretized
into 300 volume elements resulting in a cell size of $\Delta x=0.5$cm.
For the MRMT model of the complex fracture network the fracture is discretized
into 600 volume elements resulting  in a cell size of $\Delta x=2.5$cm. The memory functions
are expanded into exponential functions, making the model equivalent to one with a mobile
continuum and a number of N independent immobile single rate transfer continua.
The matrix is therefore represented by another N$\times$300 respectively N$\times$600 cells,
where N is the number of immobile boxes. For the full 2D model the domain for the simple
fracture network is discretized into 180 (perpendicular to the flow direction) $\times$ 3000 (in flow
direction) elements resulting in a cell size of $\Delta x=0.05$cm.
The domain for the complex fracture network is discretized in 445 $\times$
1422 elements resulting in a cell size of $\Delta x=1$cm.

As simulation results the wetting fluid saturation distribution from the
detailed simulation, the breakthrough curve, here defined as the averaged non-wetting saturation in the fracture at the right boundary of the observation volume and the oil recovery $R$ are analyzed. The oil recovery is the percentage of removable
oil, that is recovered from the full domain. The recovery is

\begin{equation}
R=\frac{\phi_{m}V_{m}\left(1-\bar{S}_{m}\right)+\phi_{f}V_{m}\left(1-\bar{S}_{f}\right)}{\phi_{m}V_{m}+\phi_{f}V_{f}}
=\frac{a_{v}\left(1-\bar{S}_{m}\right)+\left(1-\bar{S}_{f}\right)}{a_{v}+1}.
\end{equation}

Here $\bar{S}_{m}$ and $\bar{S}_{f}$ are the movable water saturation,
calculated as

\begin{equation}
\bar{S}_{i}(t)=\frac{\frac{1}{V_{i}}\int_{V_{i}}S_{i}(t)dV_{i}-S_{i}^{0}}{S_{i}^{max}-S_{i}^{\text{0}}}\quad i=\{f,m\}.
\end{equation}

where $V_{f}$ and $V_{m}$ are the total volumes of the fractures
respectively the total volume of the matrix blocks. In the examples
these are the volume of the fractures respectively the volume of the
matrix blocks in the observed volume as defined in figure \ref{fig:fracture-networks}.
$S_{i}^{max}$ is the maximal water saturation. In these examples the
maximal water saturation is $S_{i}^{max}=1$ in the fracture network
and in the matrix blocks.

\subsection{Simple Fracture Network: Time Scales}
In the first example, we want to show the applicability of the model
by comparing the full 2D model to the 1D MRMT model for different
ratios of viscous time scale in the fracture and the capillary diffusion time
scale in the matrix blocks. We keep the viscous timescale $t_{q,f}$
for the fracture constant and vary the timescale $t_{\omega}$ for the
capillary flow in the matrix blocks by varying the intrinsic permeability
in the matrix block.

We show four different cases as given in table
\ref{tab:Parameters-for-example-1}. For the first case (A) we have a faster flow in the matrix than in the fracture,
$t_{q,f}>t_{\omega}$ and for the second case (B) the time scales are comparable $t_{q,f}\approx t_{\omega}$.
For the third and fourth cases (C1, C2) we choose the parameters such that $t_{q,f}<t_{\omega}$,
and expect that the flow in the fracture is much faster than the
flow in the matrix.

The parameters for the MRMT model can be found from the geometry of
the fracture network and the flow parameters. We calculate $a_{v}=5.1$
for the volume ratio and $q_{t}=5.26\times10^{-5}\textrm{m/s}$ for the flow velocity.
In this study case, all local memory functions $\varphi_{\omega}$ are calculated numerically
(See \ref{num:memoryfunction} for details). For the numerical simulation,
we use a Cartesian grid with a cell size of $\Delta x=0.05\textrm{cm}$.
The global memory function is then calculated from~(\ref{eq:globalmemory2}).

\begin{figure}
\noindent \begin{centering}
\includegraphics[width=12cm]{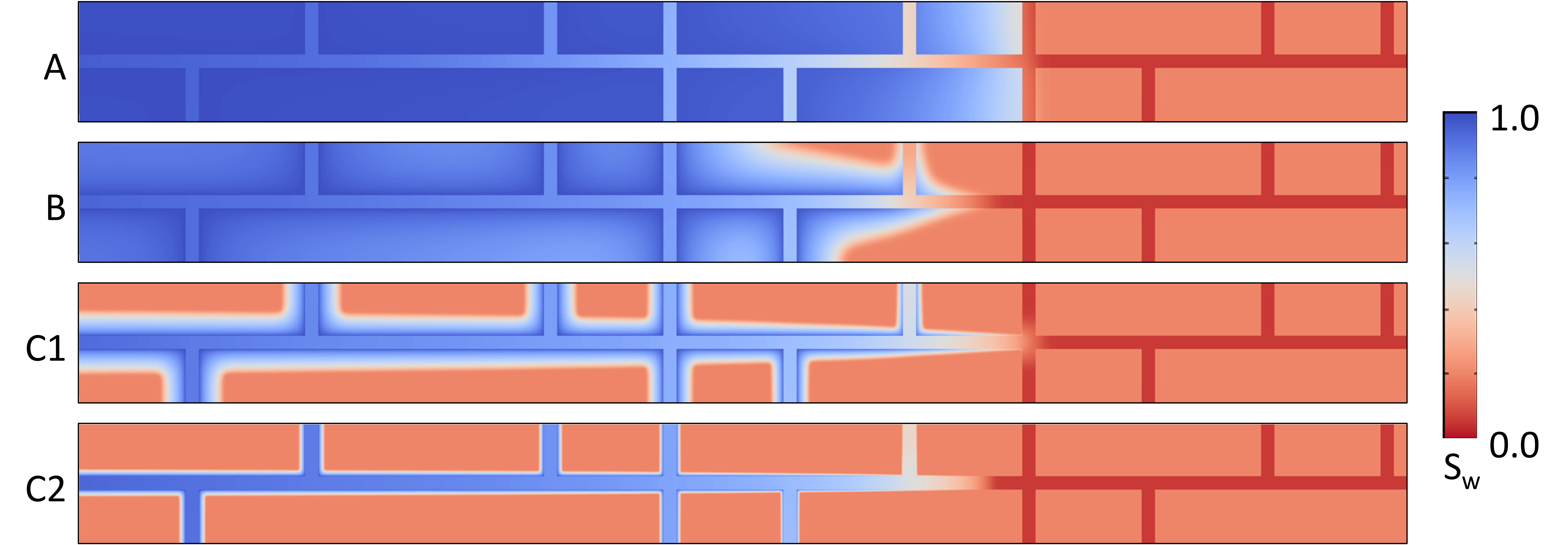}
\par\end{centering}

\protect\caption{\label{fig:Sat-profiles-fracture-50}Saturation distribution for cases
A, B, C1 and C2 with increasing capillary diffusion time scale from top
to bottom, at the time when the mass of oil in the fracture is reduced to 50\% in
the observed region compared to the initial state. This takes place
for A at $t_{f,50\%}=5.9\times10^{4}$s, for B at $t_{f,50\%}=4.7\times10^{4}$s,
for C1 at $t_{f,50\%}=2.1\times10^{4}$ and for C2 at $t_{f,50\%}=1.2\times10^{4}$s.}
\end{figure}

\begin{figure}
\noindent
\centering{}\includegraphics[width=12cm]{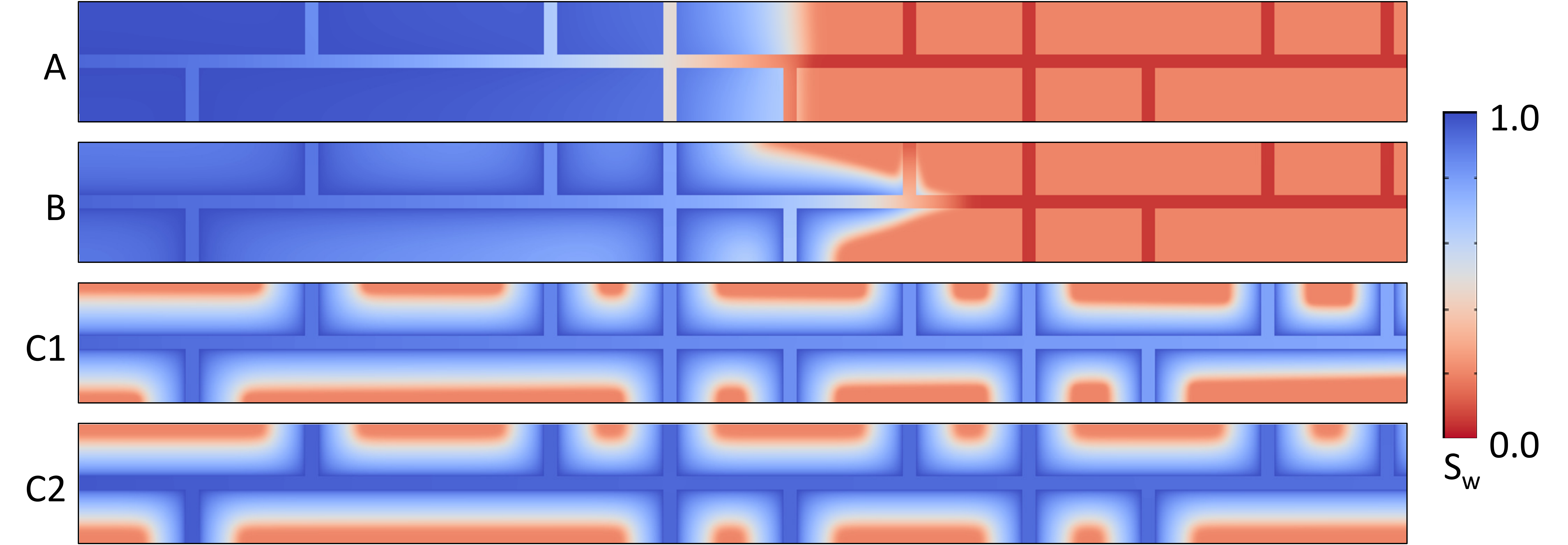}
\protect\caption{\label{fig:Sat-profiles-matrix-50} Saturation distribution
for cases A, B, C1 and C2 with increasing capillary diffusion time scale
from top to bottom, at the time when the mass of oil in the matrix is reduced to
50\% in the observed region compared to the initial state. This takes
place for A at $t_{m,50\%}=4.4\times10^{4}$s, for B at $t_{m,50\%}=4.6\times10^{4}$s,
for C1 at $t_{m,50\%}=6.8\times10^{4}$s and for C2 at $t_{m,50\%}=5.2\times10^{5}$s.}
\end{figure}

\begin{figure}
\noindent \begin{centering}
\includegraphics[width=10cm]{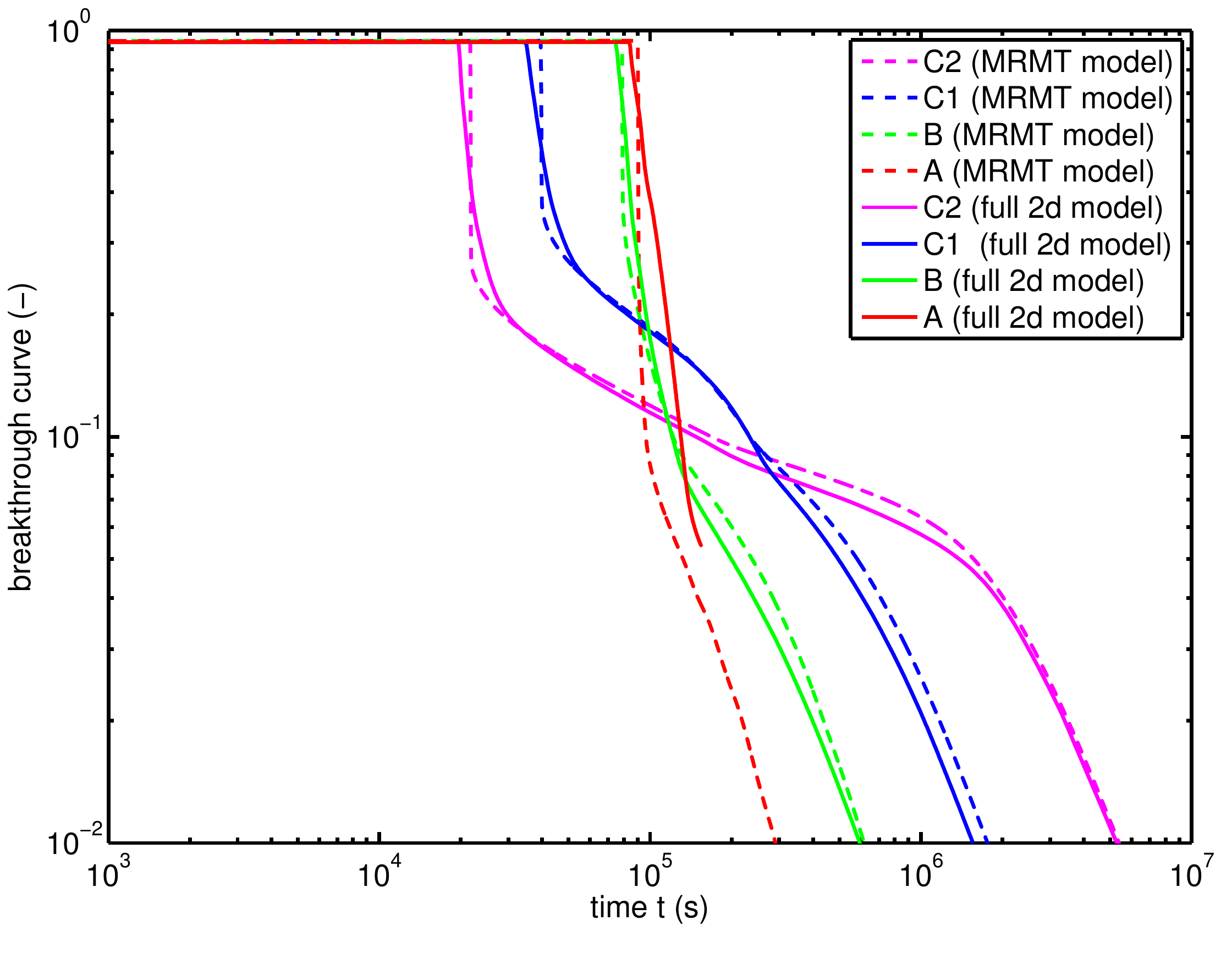}
\par\end{centering}

\noindent \begin{centering}
\includegraphics[width=10cm]{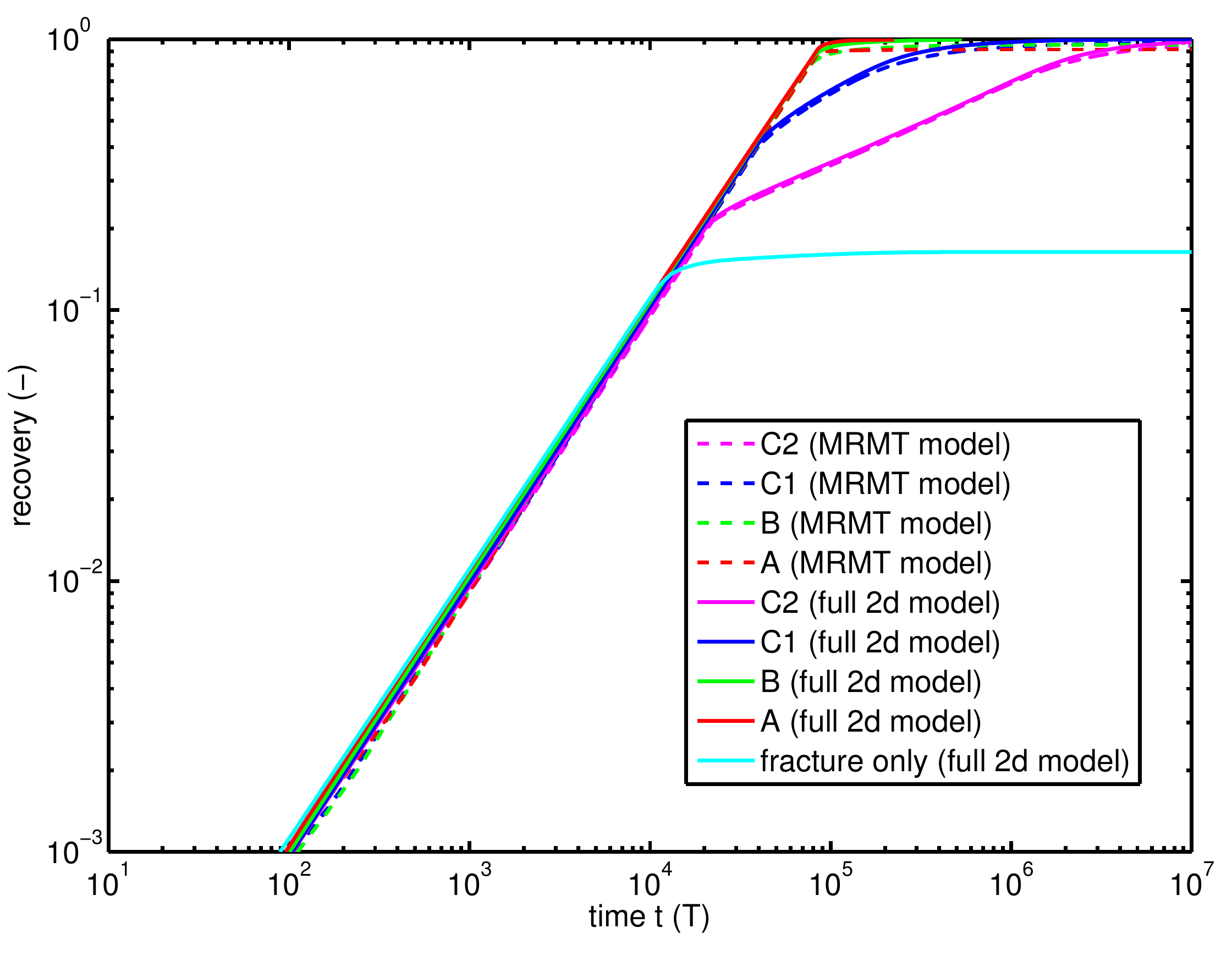}
\par\end{centering}

\protect\caption{\label{fig:results_simple}Breakthrough curve (top) and recovery (bottom) for the full 2d models
and the MRMT models for the first example.
}
\end{figure}

The saturation distributions from the full 2d models can be found
in figures \ref{fig:Sat-profiles-fracture-50} and \ref{fig:Sat-profiles-matrix-50}.
In figure \ref{fig:Sat-profiles-fracture-50} we show the saturation
distribution, when the mass of oil in the fracture is reduced to 50\%
compared to the initial state and in figure \ref{fig:Sat-profiles-matrix-50}
we show the saturation distribution, when the mass of oil in the matrix
is reduced to 50\% compared to the initial state to verify the time
scales. The results of the MRMT model and the full 2d
model can be compared in figure \ref{fig:results_simple}.

As expected from the time scales in table \ref{tab:Parameters-for-example-1}, we find two different types of flow in the fracture network (see
figures \ref{fig:Sat-profiles-fracture-50} and \ref{fig:Sat-profiles-matrix-50}).
In the cases C1 and C2 (table \ref{tab:Parameters-for-example-1}), the flow in the fracture is faster than the flow in the matrix.
The fracture fills up first and the matrix blocks fill up after the
breakthrough. It should be noted that although there are different matrix block sizes in the fracture network (see figure \ref{fig:fracture-networks}), the time scale for filling due to capillary counter current flow is almost the same for all of them. The shortest length scale of the matrix blocks is the distance to the outer boundaries, which is the same for all matrix blocks. The tailing (see figure \ref{fig:results_simple}) is caused by an interaction of matrix-fracture transfer and rarefaction in the fracture network. As the capillary diffusion time scale $t_{\omega}$ is representative for all matrix blocks, it predicts when the recovery in the full system is almost complete. The tailing in the breakthrough curves in figure \ref{fig:results_simple} have for this reason all the same slope in the log-log plot (caused by the rarefaction) and break off at different times, which correspond in the case C1 and C2 to the capillary diffusion time scale $t_{\omega}$ ($10^{5}$ s for C1 and $10^{6}$ s for C2). In case A and B this time scale would be before breakthrough time. The resulting non-equilibrium flow behavior is well represented by the presented MRMT model.

In the cases A and B (table \ref{tab:Parameters-for-example-1}), where the flow in the fracture is equally fast or slower than the flow in matrix, the displacement front in the fracture stops when it
touches a matrix block, until this matrix block is filled up and continues to move after that. This becomes apparent in figures \ref{fig:Sat-profiles-fracture-50} and \ref{fig:Sat-profiles-matrix-50}. Here the flow in the matrix
is unidirectional. When comparing the recovery curves in figure \ref{fig:results_simple} it can be seen that, despite the questionable approximation of the matrix flow as counter-current, the MRMT model still gives a good representation of the process, also for case B. The flow behavior in the fracture network for this case can be modeled with an equilibrium
model, as can clearly be observed in figures \ref{fig:Sat-profiles-fracture-50} and \ref{fig:Sat-profiles-matrix-50}, and there is no need for the MRMT model. An equilibrium model is just a simpler option. However, we would like to stress that the MRMT model, although in this case more complex than needed, does make good predictions of the averaged flow.

The timescales that are here suggested differentiate between these flow behaviors and can be used to estimate the flow regime.
The capillary diffusion time scale characterizes the flow for the matrix well
and the length scale (\ref{eq:Ma}) of Ma et al. \cite{ma1997} gives a good estimate
for the relevant length scale of the matrix blocks that is needed to estimate the time scale for counter current flow.
This capillary diffusion time scale $t_{\omega}$ can be used to parameterizes the local memory function in the matrix block and will be used in the estimation suggested in the next section.

\subsection{Complex Fracture Network: Estimating the Global Memory Function}

\begin{figure}
\noindent \begin{centering}
\includegraphics[width=10cm]{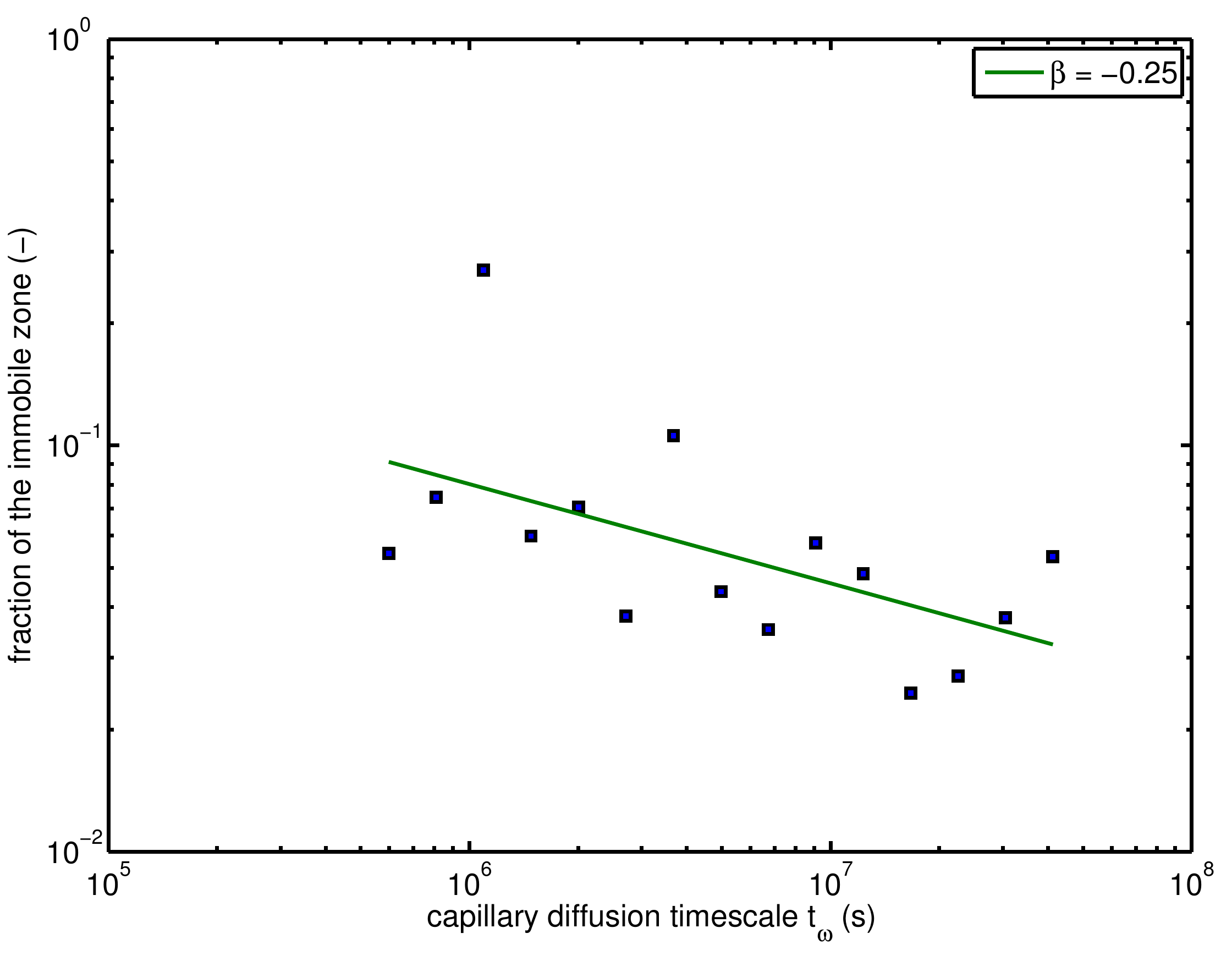}
\par\end{centering}

\protect\caption{\label{fig:Histogramm}Histogramm of the capillary diffusion time scales in
the matrix blocks for the complex fracture network}
\end{figure}

In the second example we demonstrate the scaling approaches to approximate the memory functions. The benefits of this MRMT model emerge for broad distributions of local capillary diffusion time scales. These time scales depend on different properties of the matrix blocks such as the characteristic length scale, the intrinsic permeability or the porosity. 
For the sake of computational efficiency we focus on the variability of length scales of the matrix blocks and choose a simple geometry for the fracture network. The resulting complex fracture network can be found in figure~\ref{fig:fracture-networks}. This network seems artificial, however, we would like to stress that a more realistic representation of the shapes of the matrix blocks would not lead to a very different distribution of characteristic length scales or a qualitatively different breakthrough curve (tested with a smaller test case and not shown). However, it would have been computationally too expensive, considering that the calculations shown took several months on a computer cluster. Also, one could have included a variability of permeabilities and other parameters in the matrix blocks. This leads, however, to very slow convergence of the numerical simulations and would not add anything to the distribution of capillary diffusion time scales.
The characteristic length scales in our test example vary over one order of magnitude. This results in a distribution of capillary diffusion time scales over two orders of magnitude as shown in figure~\ref{fig:Histogramm}. These capillary diffusion time scales $t_{\omega}$ are larger than the advective timescale $t_{q,f}$. In this regard this example is similar to the first test cases C1 and C2.

For this complex fracture network, it is computationally expensive to
calculate all local memory functions numerically. Therefore
we demonstrate the two different methods introduced before. For
the first method we calculate one local memory function for a reference
block $\omega_0$ numerically (see \ref{num:memoryfunction} for details).
Then we approximate all other local memory functions
by scaling this reference memory function $\varphi_{\omega_0}$ with the associated characteristic length scales of the matrix blocks.
We evaluate this by numerical interpolation. For the characteristic length (\ref{eq:Mablock}), we estimate the
length and width of each block in the fracture network with image analysis.

For the second
method, we approximate the local memory functions by a truncated
power law, which is given by (\ref{eq:truncated_power_law}) and (\ref{scaling}) as
\begin{equation}
\varphi_{\omega}(t)=\frac{1}{t_{\omega}}\frac{\exp(-t/t_{\omega})}{\Gamma(1/2)\sqrt{t/t_{\omega}}}
\end{equation}
Here we calculate the McWorther
solution once to calculate the capillary diffusion time scales $t_{\omega}$
in the matrix blocks, but we do not need to do any numerical simulations
on 2d grids to find the local memory functions.

For both methods, the global memory function is calculated with~\eqref{eq:globalmemory2}.
The other parameters for the MRMT model can be found from the geometry
of the fracture network and the flow parameters. The volume ratio
is $a_{v}=1.84$ and the mean total Darcy velocity is $q_{t}=2.3\times10^{-4}$m/s.

For comparison we also set up the MRMT model with a memory function where only one local memory function derived with 
a single capillary diffusion time scale $\left\langle t_{d}\right\rangle$ is used as global memory function. This means, no superposition is carried out, but the assumption is made that one capillary diffusion time scale is representative for the whole matrix domain. For this we use the averaged capillary diffusion time scale, $\left\langle t_{d}\right\rangle =\sum_{\omega} n_{\omega} \chi_{\omega}t_{d,\omega}=3.7\times10^{6}$s (cf. figure \ref{fig:Histogramm}). The associated memory function is calculated numerically.

To summarize the parameterization: The parameters for the MRMT model are calculated from fluid and material
properties and the structure of the fracture network. The memory
function is the crucial parameter to capture the fracture matrix exchange. It is calculated for the fracture network
from the distribution (see figure \ref{fig:Histogramm}) of matrix block sizes and the hydraulic parameters
of the matrix blocks. Other details of the structure, like the location
of each block in the fracture network, are not important, because
the flow in the fracture is faster than the flow in the matrix.

\begin{figure}
\noindent \begin{centering}
\includegraphics[width=12cm]{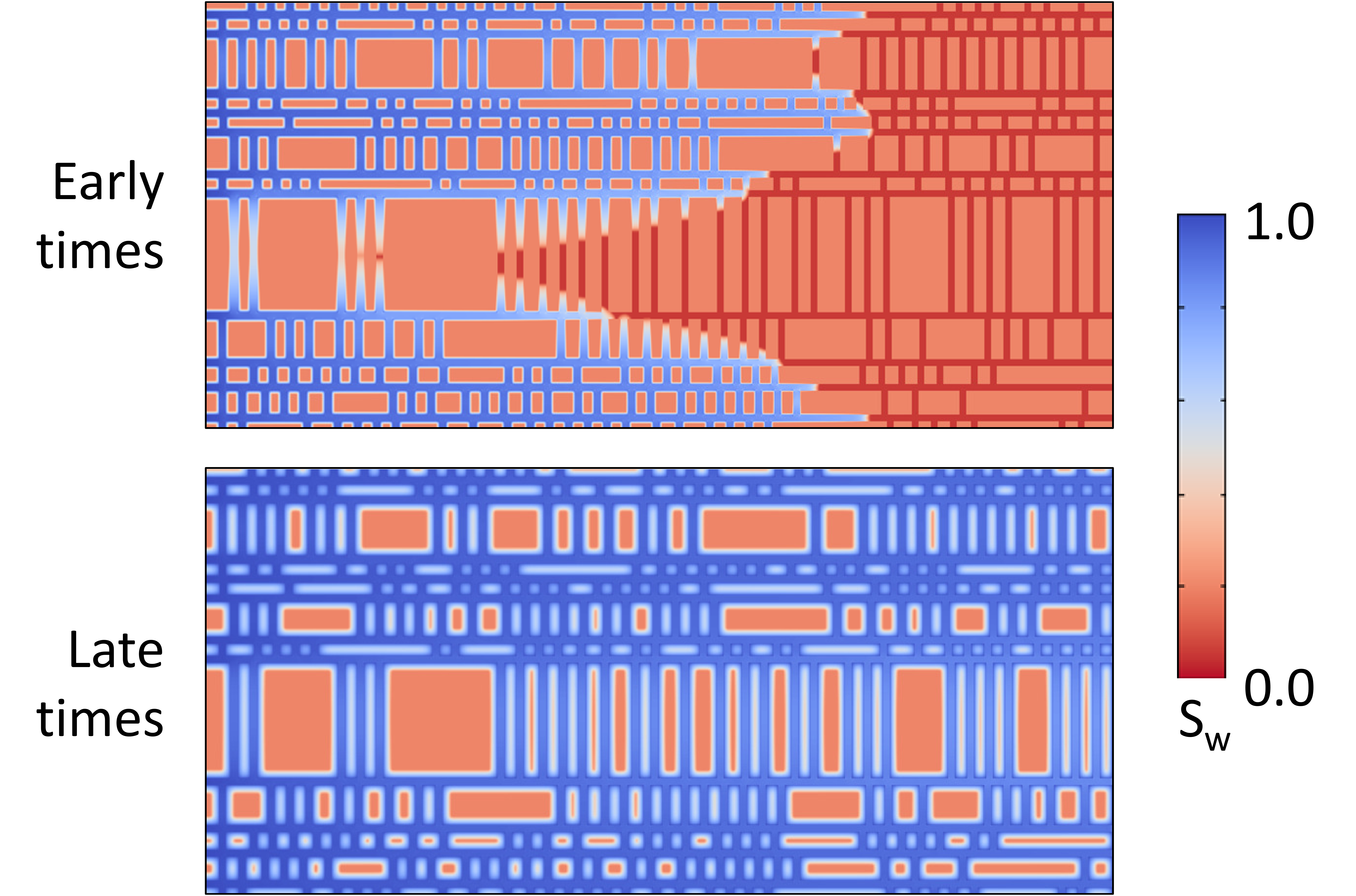}
\par\end{centering}

\protect\caption{\label{fig:Saturation-profiles-complex}Saturation distribution obtained with the
full 2d model before (top) and after (bottom) breakthrough for the
complex fracture network. }
\end{figure}

\begin{figure}
\noindent \begin{centering}
\includegraphics[width=10cm]{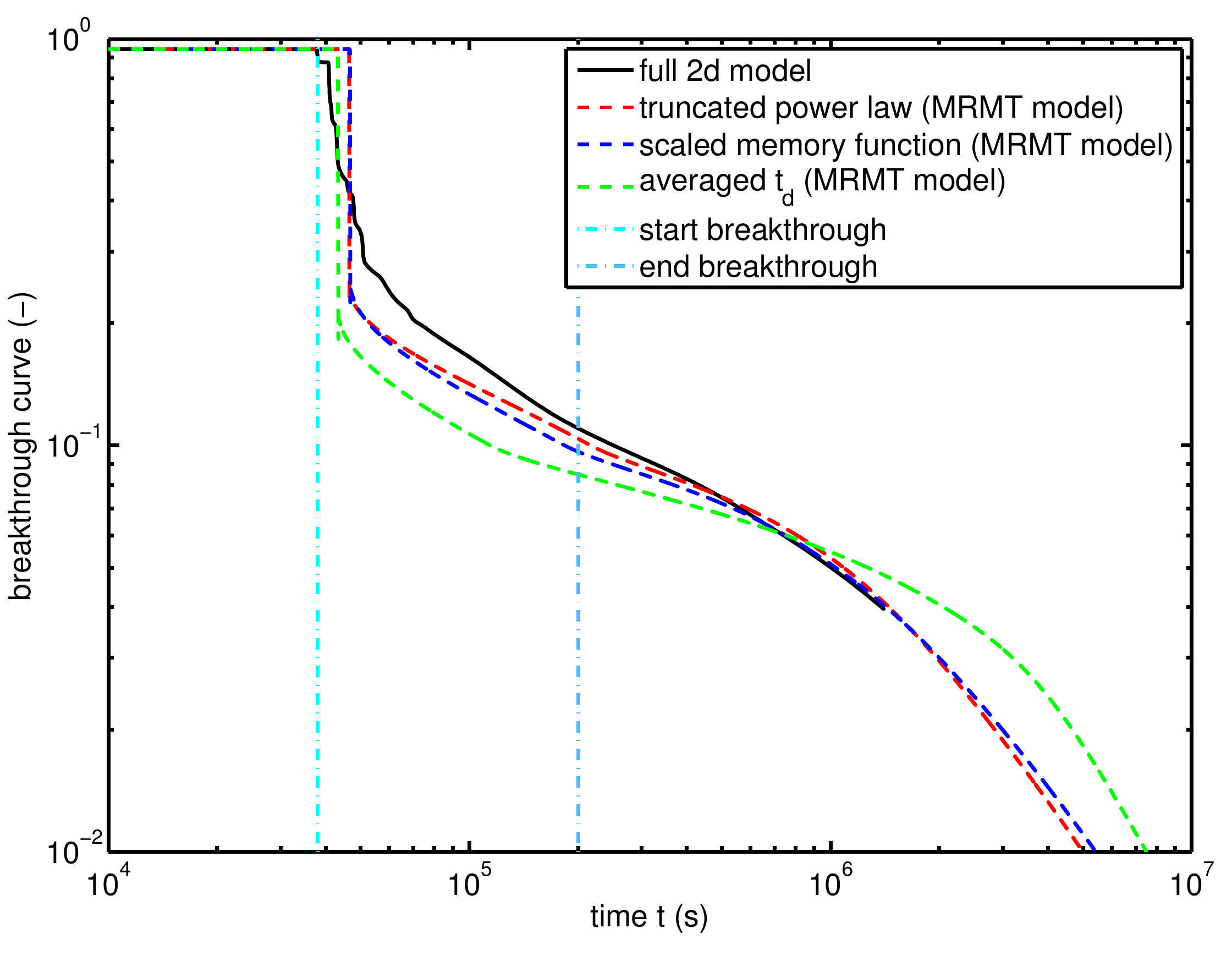}
\par\end{centering}

\noindent \begin{centering}
\includegraphics[width=10cm]{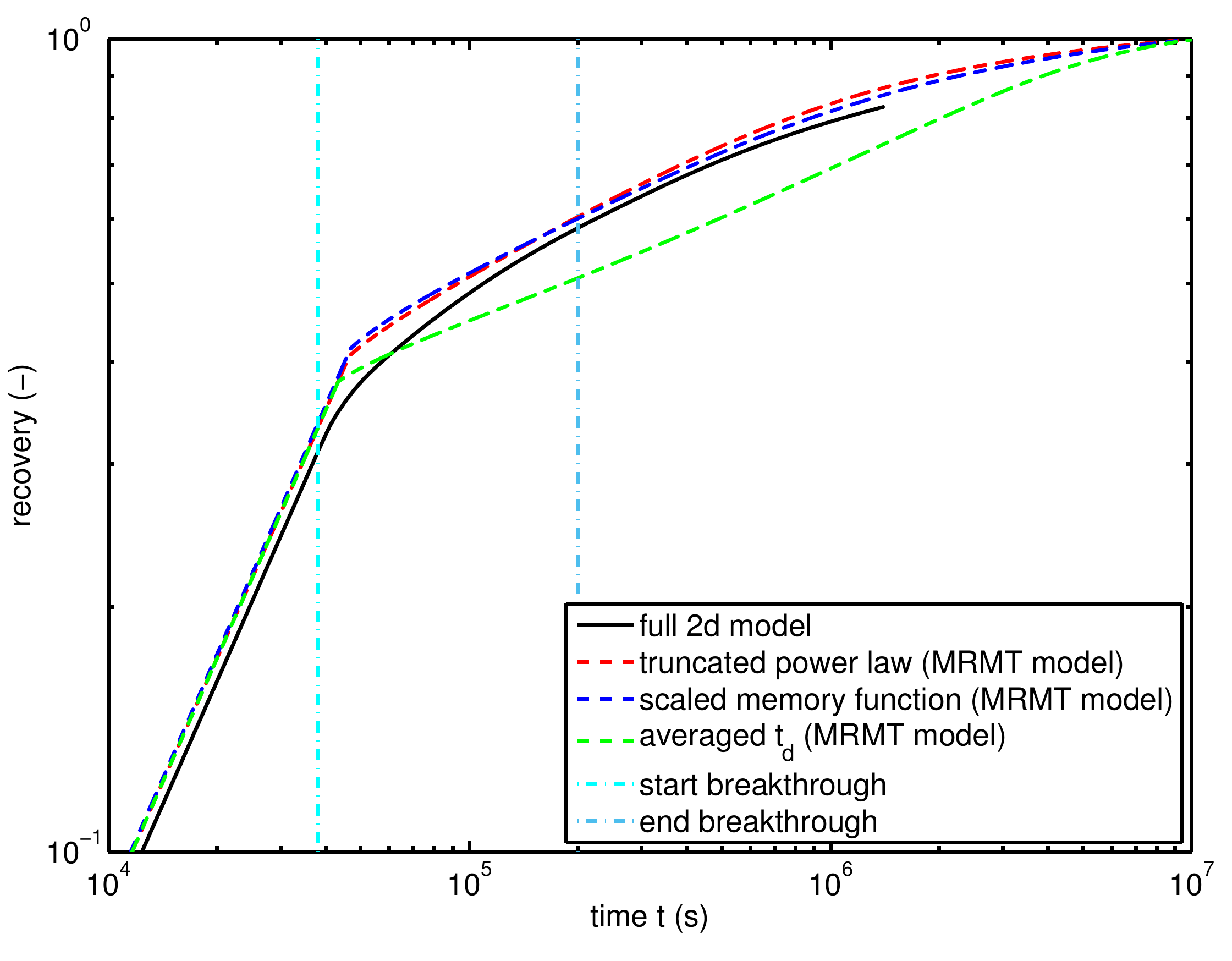}
\par\end{centering}

\protect\caption{\label{fig:result_complex}Breakthrough curve
(top) and recovery (bottom) for the full 2d model and the MRMT
model for the complex fracture network.}
\end{figure}

All memory functions for this example are shown in figure~\ref{fig:memoryfunctions}.
The saturation distribution from the full 2d model for a time before and
after breakthrough can be found in figure \ref{fig:Saturation-profiles-complex}.
The results obtained with the MRMT model with approximated global memory functions
and the full 2d model can be compared in figure \ref{fig:result_complex}.

The MRMT models that consider the full distribution of time scales matches the recovery curves as well as the breakthrough curves of the detailed model well (see figures~\ref{fig:results_simple} and \ref{fig:result_complex}). In particular the averaged breakthrough time and the late time behavior can be reproduced by this MRMT models. There are some differences around the breakthrough time. The recovery curve as well as the breakthrough curve of the detailed model are spread around the breakthrough time. This has two reasons: (i) the detailed model considers capillary diffusion in the fractures. Furthermore (ii) the saturation front speed in the detailed model varies locally. This can be seen clearly in figure~\ref{fig:Saturation-profiles-complex}. This is caused by local variations in the fracture geometry and in the retardation due to fracture matrix exchange. Neither capillary diffusion nor a distribution of front speed are captured in the MRMT model, so that the MRMT model simulates a mean breakthrough.

The tailing of the breakthrough curve (see figure \ref{fig:result_complex}) is caused by rarefaction in the fracture domain and by the superposition of the different breakthrough times. In comparison with the MRMT model with the single averaged capillary diffusion time scale, the tailing of the breakthrough curve shows a different slope. This different slope is caused by the distribution of capillary diffusion time scales (figure~\ref{fig:Histogramm}), which is not captured when using only one equivalent local memory function. Because the global memory functions calculated from the truncated power law and the numerical simulations differ at late time (see figure~\ref{fig:memoryfunctions}) it is expected, that the memory function derived from the numerical simulation is more accurate.

The breakthrough time depends on flow rate at the inlet and on the matrix-fracture transfer. The flow rate at the inlet is the same for all MRMT models. The MRMT model with the single averaged capillary diffusion time scale shows an earlier breakthrough than the MRMT model with the full global memory function, because this model ignores the fast capillary diffusion time scales. Therefore this model underestimates the matrix-fracture transfer at early times. This can be read from the memory functions in figure~\ref{fig:memoryfunctions} directly.

Overall the MRMT model with the averaged capillary diffusion timescale results in earlier breakthrough and does not match the tailing of the
full 2d model. The solution of the one dimensional MRMT model that capture the full distribution of capillary diffusion time scales matches the solution of the full two dimensional model for the complex fracture network, despite the simplifications made. This illustrates that all capillary diffusion time scales need to be considered in a case of a complex fracture network.

\section{Summary and Conclusions}
We show the application of a MRMT model with pre-estimated parameter
functions to simulate immiscible displacement in complex fracture networks.
In order to estimate the applicability of the MRMT approach, we
introduced time scales to characterize the flow in fracture and matrix.
The presented model is valid for imbibition cases, when the flow in the fracture
is faster than the flow in the matrix and the fracture network is
sufficiently well connected. The parameters for the MRMT model can
be predicted from fluid and material properties and the structure of the fracture network.
The global memory function can be calculated from the weighted
sum of volume percentage of the local memory functions for each matrix block
in the fracture network. The local memory function for a single block can
be calculated numerically or it could be approximated or by using a truncated power law parameterization using the capillary
diffusion timescale as break off time. It is practical to approximate
the local memory functions for other matrix blocks by scaling a reference
memory function.  

The MRMT model presented here can be transformed to and solved
as MRDP model. Therefore, it is easy to implement this model in 
existing MRDP simulators. It would only be necessary to expand the
pre-estimated global memory function into exponential functions (see
Appendix \ref{implmrmt}). The MRDP medium would need as many continua
as there are expansion terms for the memory function. The parameters
could be read directly from the expansion. However, the model is based
on assumptions that should be pointed out clearly.

We assume that the structure of the fractured rock is known. The information
about the block size distribution and the hydraulic properties of all materials
are needed to calculate the parameters for the MRMT model. The hydraulic
parameters and the reference memory function for a real setup could
be obtained from laboratory experiments using samples from bore cores,
for example. Information about the structure
of fracture network is more difficult to obtain and could be available
from outcrops. A reconstruction of
the fracture network from outcrops will only be an estimation. It
should, however, be noted that only the distribution of block sizes
is needed to calculate the global memory function, while details about the
locations of the matrix blocks are not required.
Furthermore, the definition of the matrix block sizes may be difficult
for real three dimensional fracture networks. The topology of
two-dimensional networks is fundamentally different from that of
three-dimensional networks, where matrix blocks might not be
completely surrounded by fractures.

A fundamental assumption of the upscaled multi-rate mass transfer
model is that the fracture network itself can be represented by a
continuum with effective properties. The fracture network is here
represented as a fairly homogeneous continuum with a straightforward
way to estimate effective parameters because the focus of this paper
is to capture the fracture-matrix exchange in an upscaled model.
If the fracture network is complex with lots of dead ends and not well-defined matrix blocks,
the approximation of the global memory function from the superposition of scaled
blocks is expected to become more complex or to fail. The fracture
network needs to be sufficiently well connected. This is related to
the requirement that a macroscopic representative elementary volume
needs to exist for the medium. If this is not fulfilled, the continuum
approach for the fracture network fails and it may be better to apply
the DFM approach.

If the fracture network is heterogeneous, but representable as a
continuum, the upscaled model for the fracture needs to be
extended by effective parameters and possibly by effective
processes. In the complex fracture network, we found a spreading of
the front caused by the heterogeneity of the fracture network itself. This front
spreading may be captured by an additional dispersion term
for flow in the fracture network~\cite[][]{langlo, neuweileretal03,
  BDC2009}.  

Strong fracture network heterogeneity and long rage structures, as
caused, for instance by variation in the fracture aperture, may cause preferential
flow paths. Such a scenario might happen if
there are few intersecting fractures with large aperture in a network
with smaller aperture fractures. This might lead to a fast
breakthrough along these preferential paths, while other parts of the
fracture network do not contribute to the displacement process. The
capillary flow within these passive fractures will be much weaker than that into
the matrix blocks. In this case, the fracture network would need be
represented by a more advanced upscaled continuum model. A possible
approach would be so separate the fracture network itself into a
mobile continuum, containing the contributing fractures, and assign
non-participating fractures to a second immobile continuum with little
exchange or even represent it with a reduced porosity of the
medium. The topology of these non-participating fractures will
influence the global memory function, as they will form barriers in
the matrix blocks. The pre-estimation of these continua for a given
fracture network is not straight forward and goes beyond the scope of
this work. We would like to stress that this applies to fracture networks with
long range structures only. If the apertures of the fractures are
highly heterogeneous, but the correlation scale of aperture
variability is much smaller than the intersecting length of fractures,
the flow parameters of the fracture might need to be represented by
effective parameters. 

Apart from the fracture network and the fracture apertures, 
the surrounding matrix is usually also influenced by heterogeneous material structure.
 This is not addressed in this work, but could be included straight away 
for the case that heterogeneity is moderate, meaning the variance of parameter contrasts
 is low and the correlation length of heterogeneity is small.  If the matrix is
 moderately heterogeneous, the two-phase flow parameters in the matrix can be considered
 effective parameters, describing the spatial average distribution. 
For slow flow that is locally in equilibrium, such as for capillary counter current flow,
 this is a reasonable approach. If heterogeneity is strong, the matrix 
cannot be described well as one continuum. In this case the MINC approach
 \cite{pruess1992brief} could be suitable.

A comparable effect causing preferential flow paths is segregation of
fluids due to buoyancy. Gravity is not considered in this model,
but will influence the flow in the fractured rock. Gravity affects
the flow in the fracture network more than the flow in the matrix, as
the gravity number, which compares gravity to capillary forces
\cite{HilferandOren96}, is proportional to the permeability of the
medium. If the injection of fluid is horizontal (for example in a deep
aquifer layer), gravity will lead to a flow segregation with the
denser fluid flowing in the fractures at the bottom of the layer
\cite[e.g.,][]{Dietz1953}). The approach outlined in this paper could
be used in a straight forward manner, but would require a two or three
dimensional extension of the MRMT model, which is straightforward in
the presented framework. 

\section{Acknowledgements }
This work was supported by the compute cluster, which is funded by
the Leibniz University Hanover, the Lower Saxony Ministry of Science
and Culture (MWK) and by the German Research Association (DFG) under
the grant NE 824 10-1. We gratefully acknowledge the help of Bernd
Flemisch from the University of Stuttgart with the Dumux
model. M.D. acknowledges the funding from the European Research
Council through the project MHetScale (Grant agreement no. 617511).

\appendix

\section{Numerical Implementation of the MRMT Model}\label{implmrmt}
The MRMT model~\eqref{upscaled:model} is numerically
solved as presented in Silva et al. \cite{Silva2009} or Tecklenburg et al. \cite{tecklenburg2013}. Other schemes might be possible, such as suggested by Di Donato et al. \cite{DiDonato2007}.

To avoid carrying out an integration over time in the numerical scheme, which requires a very large amount of storage, the
convolution integral modeling the sink source term is approximated
by a sum of single rate transfer functions following \cite{Willmann2010}.
Hence, the global memory function can be approximated by a sum of $N$ exponential functions.

\begin{equation}
\varphi(t)=\sum_{j=1}^{N}\alpha_{j}b_{j}\exp\left(-\alpha_{j}t\right)
\end{equation}

Here $b_{j}$ (-) are weighting coefficients and $\alpha_{j}$ ($\mathrm{T}^{-1}$)
are the transfer rates. The exponential functions can be considered the solutions
of the mass balance in immobile boxes with single rate transfer with
the fracture domain.

\begin{equation}
T_{j}=b_{j}\frac{\partial S_{j}}{\partial t}=b_{j}\alpha_{j}\left(S_{mb}-S_{j}\right)\label{eq:singlerate}
\end{equation}

The convolution integral in~\eqref{upscaled:model} can with this approach be approximated
by a weighted sum of single rate transfer functions with $N$ immobile
boxes (see \cite{Willmann2010} for details).

\begin{equation}
T=\frac{\partial}{\partial t}\int_{0}^{t}\varphi(t-t')S_{mb}(t')\mathrm{d}t'+S_{w,m}^{0}\varphi(t)\approx\sum_{j=1}^{N}T_{j}=\sum_{j=1}^{N}b_{j}\alpha_{j}\left(S_{mb}-S_{j}\right)\label{eq:equivalence}
\end{equation}

$S_{j}$ (-) is the saturation in the immobile box $j$, We then find
the weights $w_{j}=b_{j}\alpha_{j}$ by minimizing the functional

\begin{equation}
\mathcal{F}=\sum_{k=1}^{M}\sum_{j=1}^{N}\left\{ w_{j}\exp\left(-\alpha_{j}t_{k}\right)-\varphi\left(t_{k}\right)\right\} ^{2}
\end{equation}

For the logarithms of the times, $1/\alpha_{j}$, we choose evenly distributed
values between $\ln(1/\alpha_{1})$ and $\ln(1/\alpha_{N})$. For
$1/\alpha_{1}$ we choose a fraction of the time step size $\Delta t/10$
and for $1/\alpha_{N}$ the maximum simulation time $t_{end}$. In
the simulation we use five exponential functions per decade. Here
$t_{1}$ is the first and $t_{M}$ is the last time step of the simulation.

With the approximation of the kernel of the convolution integral by a sum of exponential
functions, we can now solve the MRMT~\eqref{upscaled:model} following \cite{Silva2009}.
We introduce a one dimensional grid with volumes $i$ to discretize the upscaled fracture
network. Each volume $i$ is connected to $N$ immobile boxes $j$ representing the matrix.
The initial fracture saturation is $S_{i}^{0}=S_{f,m}^{0}$ and the initial matrix saturation
is $S_{i,j}^{0}=P{}_{c,m}^{-1}\left\{ P_{c,f}(S_{f,m}^{0})\right\} =S_{w,m}^{0}$
for all boxes $j$ and volumes $i$.

We apply an operator splitting scheme: First the exchange between
matrix and fracture is evaluated, then the flow in the fracture is
solved by an upwind finite volume method and explicit Euler time integration
as described in \cite{leveque1992numerical}.

Assuming $S_{mb}=S_{mb}^{k}$ to be constant between time steps $k$
and $k+1$, each single rate transfer (\ref{eq:singlerate}) can be
solved analytically using $S_{j}^{k}$ as the initial condition.

\begin{equation}
S_{i,j}\left(t\right)=S_{mb,i}^{k}\left(1-\exp\left(-\alpha_{j}\left(t-t_{k}\right)\right)\right)+S_{i,j}^{k}\exp\left(-\alpha_{j}\left(t-t_{k}\right)\right)\label{eq:single_analytisch}
\end{equation}

The exchange between the fracture and the immobile box $j$ can be
found by substituting (\ref{eq:single_analytisch}) into (\ref{eq:singlerate})
and with defining $\Delta t=t_{k\text{+}1}-t_{k}$:

\begin{equation}
T_{i,j}^{k}=w_{j}\left(S_{mb,i}^{k}-S_{i,j}^{k}\right)\exp\left(-\alpha_{j}\Delta t\right)\quad\forall i,j.
\end{equation}

We calculate the exchange between fracture and matrix to get the intermediate
fracture saturation $S_{f}^{int}$

\begin{equation}
S_{f,i}^{int}=S_{f,i}^{k}-a_{v}\sum_{j=1}^{N}T_{j}^{k}\quad\forall i.
\end{equation}

From this intermediate fracture saturation $S_{f}^{int}$ we calculate
the advective flow in the fracture using upwind fluxes. When $q_{t,f}>0$,
the fracture saturation is

\begin{equation}
S_{f,i}^{k+1}=S_{f,i}^{int}-\frac{\Delta t}{\Delta x}\cdot\frac{q_{t,f}}{n_{f,f}}\left\{ f_{f}\left(S_{f,i}^{n-1}\right)-f_{f}\left(S_{f,i-1}^{n-1}\right)\right\} \quad\forall i.
\end{equation}

Here the time step size is controlled by the Courant Friedrich Levy
Criterion.

The immobile boxes are updated using (\ref{eq:single_analytisch})
as $S_{i,j}^{k+1}=S_{i,j}\left(t_{k+1}\right)$ and then the averaged
saturation in the matrix can be calculated as

\begin{equation}
S_{m,i}^{k+1}=\sum_{j=1}^{N}b_{j}S_{i,j}^{k+1}.
\end{equation}

\section{Numerical Determination of the Local Memory Function\label{num:memoryfunction}}

The Green's function or memory function is defined as a response of
the system to a delta pulse. This is equivalent to the time derivative
of a response to a Heaviside step function. With this equivalent the
local memory function can be determined numerically.


In a numerical simulation it is more practical to apply a Heaviside
stepfunction $\Theta(t)$ as boundary condition than a delta pulse.
This step function from the initial saturation $S_{\omega}^{0}$ to
the saturation $S_{\omega,b}$ approximates the matrix saturation
at the interface between fracture and matrix during a flooding of
the fracture network. The matrix saturation is $S_{\omega,b}=P_{c,\omega}^{-1}\left\{ P_{c,f}\left(S_{f}^{0}\right)\right\} =S_{\omega}^{0}$
before the front arrives. When the front arrives the interface saturation immediately
jumps to $S_{\omega,b}=P_{c,\omega}^{-1}\left\{ P_{c,f}\left(S_{f}^{0}\right)\right\} =1$
due to the strong contrast in capillary entry pressures between the fracture
and the matrix.

The saturation $S_{\omega}(\mathbf{x},t)$ (representing the response to a Heavyside step) was here obtained with a Dumux
simulation by solving (\ref{full:model}) on a single matrix geometry. Therefore
we use the parameters for the matrix block from the detailed problem.
The initial condition is $S_{\omega}^{0}$ and the boundary condition
is $S_{\omega,b}=1$ as discussed above. We calculate the spatially
averaged saturation $\left\langle S_{\omega}(t)\right\rangle $ for
each timestep. Then the memory function can be calculated
from (\ref{Smav})
as

\begin{equation}
\varphi_{\omega}(t)=\frac{\partial}{\partial t}\left(\frac{\left\langle S_{\omega}(t)\right\rangle -S_{\omega}^{0}}{1-S_{\omega}^{0}}\right),
\end{equation}
where the time derivative is carried out by a numerical approximation, here a finite difference.

The memory function is normalized, as $\int_{0}^{\infty}\varphi_{\omega}(t)\mathrm{d}t\equiv1$.

\section{Comparison of Capillary Diffusion Time Scales\label{comparison:timescales}}

\begin{figure}
\noindent \begin{centering}
\includegraphics[width=12cm]{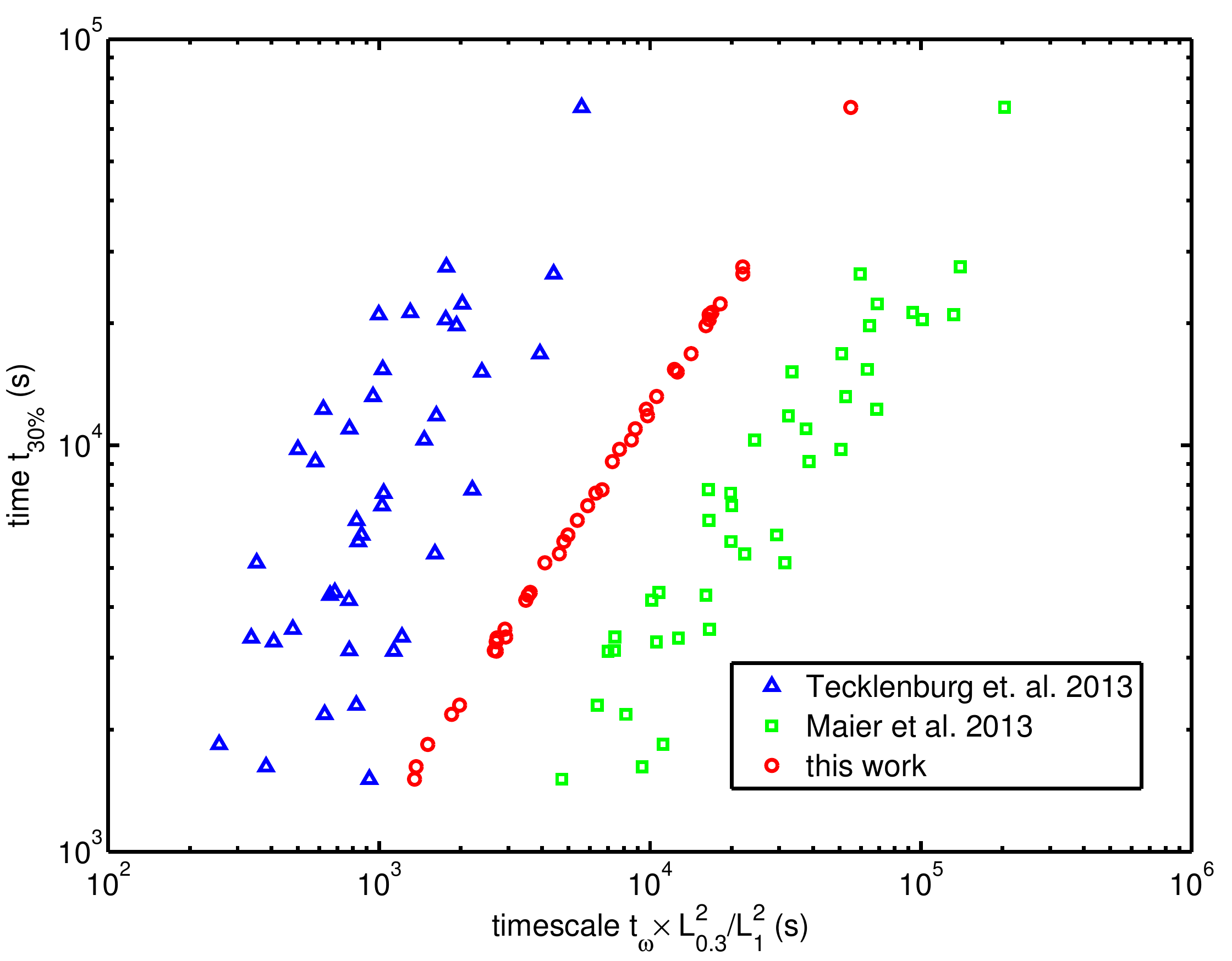}
\par\end{centering}
\protect\caption{\label{fig:Comparison-of_td}Comparison of capillary diffusion time scales
$t_{\omega}$ for imbibition into rectangular matrix blocks. $t_{30
}$ is the time, where the recovery is 30\% to consider unperturbed capillary flow only.}
\end{figure}
We now compare the new capillary diffusion time scale (\ref{eq:td_thiswork}) to other capillary diffusion
time scales from the literature that are used to parameterize MRMT or MRDP models. In these works, in contrast to our approach, the breakoff due to the interference of the displacement front with the matrix block boundaries is considered relevant. We focus on time scales based on the analytical solution of McWorther and Sunada \cite{ref:mcworther_1990},
because Schmid and Geiger~\cite{schmid2013universal, ref:schmid_2012}
compared different capillary time scales and rated this as the best
approach. The MRDP model of Maier et al. \cite{maier2013} uses a timescale
\begin{equation}
t_{\omega}=\left[\frac{L_{c}n_{f}}{2A}\right]^{2},\label{eq:tMaier}
\end{equation}
that represents the time, when the volume of water imbibed is equal to the pore volume of a slab of thickness $L_c$. The MRMT model of Tecklenburg et al. \cite{tecklenburg2013} uses the timescale
\begin{equation}
t_{\omega}=\left[\frac{L_{c}n_{f}}{2AF'(S_{0})}\right]^{2},\label{eq:tTecklenburg2013}
\end{equation}
which corresponds to the time when the front invading into a slab arrives at the boundary.

These capillary diffusion timescales are compared with the numerical solution of~\eqref{full:model} for rectangular
matrix blocks using Dumux. We simulate 40 scenarios with a boundary condition
of $S_{w,m}=1$ at all interfaces, an intrinsic permeability $K=10^{-15}\mathrm{m}^{2}$
and a grid size of $\Delta x=6.25\times10^{-5}\mathrm{m}$.
The following parameters
were chosen for a van Genuchten parameterization from the given interval by a random hypercube algorithm:
block length $a=(0.01,\,0.07)\mathrm{m}$ and width $b=(0.01,\,0.07)\mathrm{m}$,
van Genuchten $n=(1.5,\,4.5)$ and $\alpha=(1/100,\,1/1000)\mathrm{Pa}$
and the initial saturation $S_{w,m}^{0}=(0.1,0.5)$.

In these simulations one can observe an early time behavior, where the recovery is diffusive
with $\propto\sqrt{t}$ scaling, and a late time behavior, when the recovery
slows down until the matrix block is full (not shown). As we consider the early time behavior as the crucial one, we compare here the time where the recovery in the numerical simulation is 30\% and compare this (due to the diffusive time behaviour of solution ) to $t_{30\%} / t_{100\%} = L_{0.3}^2 / L_{1}^2 = 0.3^2 = 0.09$ times the capillary diffusion time scales $t_{\omega}$.

In figure \ref{fig:Comparison-of_td} we compare the timescale $t_{30\%}$
to 0.09 times the analytical capillary diffusion timescales (\ref{eq:tTecklenburg2013}), (\ref{eq:tMaier}) and (\ref{eq:td_thiswork}). We choose the timescale (\ref{eq:td_thiswork})
to characterize the flow in the matrix blocks, because this timescale
matches the early counter current flow behavior better than the other timescales.

\section*{References}
\bibliographystyle{elsarticle-num}

\end{document}